\documentclass[twocolumn,showpacs,preprintnumbers,amsmath,amssymb,floatfix]{revtex4}
\usepackage{graphicx}
\usepackage{dcolumn}
\usepackage[latin1]{inputenc}

\begin{document}

%\preprint{ArXiv}

\title {Theory of finite strain superplasticity}

\author{Miguel Lagos}
\email{mlagos@utalca.cl}
\affiliation{Facultad de Ingenier\'\i a, Universidad de Talca,
Campus Los Niches, Curic\'o, Chile}

\author{C\'esar Retamal}
\affiliation{Facultad de Ingenier\'\i a, Universidad de Talca,
Campus Los Niches, Curic\'o, Chile}
\affiliation{Universit\'e de Paris X--Nanterre, 200 Avenue de la
Republique, 9200 NANTERRE Cedex, France}

\date{November 14, 2009}

\begin{abstract}
The plastic flow of a polycrystal is analyzed assuming grains as fine
that the rate limiting process is grain boundary sliding, and grains
readily accommodate their shapes by slip to preserve spatial
continuity. It is shown that thinking of a polycrystal with randomly
oriented grains as an homogeneous and isotropic continuum when dealing
with it as a dynamical medium, even in a scale much larger than the
grain size, leads to gross errors. The polyhedral nature of grains
influences the plastic flow in a radical manner, as the relative
velocity of adjacent grains is constrained to the common boundary
plane, and only the in--plane shear stress contributes to their
relative motion. This constriction determines that $\nabla\cdot\vec
v\ne 0$, where $\vec v$ is the velocity field of the material medium,
and plastic deformation necessarily involves grain volume variations,
which can only be elastic. As this has a limit, fracture follows as a
necessary step of plastic flow. A theoretical approach to plastic flow
is developed, and emphasis is done in superplastic deformation, from
zero to fracture strain. The theory allows to quantitatively explain
the observed features of superplastic materials and responds to many
open questions in the field.
\end{abstract}

\pacs{62.20.fg, 46.35.+z, 62.20.mj, 62.20.mm}
\keywords{superplasticity, plasticity, brittleness, ductility,
fracture, grain boundaries}

\maketitle

\section{Introduction}
\label{introd}

Superplasticity is an anomalously high tensile ductility occurring in
a growing class of fine grained polycrystalline solids. At rather
precise conditions of composition, grain refinement, temperature and
strain rate, they can undergo uniform neckless plastic elongations
that may reach thousands of percent prior to fracture. These large
strains occur with no significant change in the grain shapes or sizes,
nor in any other fine attribute \cite{Pilling&Ridley, Nieh}. As well,
different bodies in the superplastic regime of deformation become
prone to diffuse one into the other when their surfaces make contact
(diffusion bonding). Among the many singular features of the
phenomenon is the fairly narrow range of temperatures over which it
happens. For instance, tensile tests for the superplastic titanium
alloy Ti--6Al--4V give maximum ductility at temperatures close to 1153
K, and elongation to fracture rapidly decays to half the optimal
figure within a neighborhood not wider than 200 K around
\cite{Cope2}. These temperatures are well below the melting point
(1930 K), so we are talking about a genuine solid state effect.

Beside the scientific problem it poses, superplasticity has acquired
considerable technical importance. After the pioneer work of Backofen
on the superplastic forming of Zn-Al sheets by a simple air pressure
operation \cite{Backofen}, as in the fabrication of plastic bottles,
and the discovery of a number of high strength superplastic alloys,
the subject moved into the scope of engineering materials for advanced
applications. Since then, the list of high strength superplastic
alloys, intermetallic compounds and ceramics has increased notably
\cite{Nieh}, together with the technical knowledge for producing
superplastic structures. A number of present day aluminium
superplastic alloys, like Al-7475, Al-8090 and Supral 220, or titanium
Ti-6Al-4V and SP700, have outstanding importance as aeronautic
materials \cite{AerospaceMat}. Superplastic forming combined with
diffusion bonding is now a standard technique for producing both
engine and airplane structural parts. The development of light ductile
materials of high strength at elevated temperatures has a principal
role in the race for more efficient aeroengines. Superplastic titanium
and nickel aluminides are drawing great attention in this respect
\cite{Nieh, AerospaceMat}. Having half the density of the current
materials, TiAl-based alloys are promising candidates to replace the
nickel-based superalloys in combustion chambers and turbine blades.

After a large plastic deformation with no significant microstructural
changes, and resembling a steady flow, superplastic solids collapse at
a rather well-defined strain, which depends on temperature and strain
rate. The regular behaviour of superplastic elongation to fracture
suggests that the phenomenon is governed by a specific variable, not
identified so far, which keeps account for the total strain and
produces failure when reaching a precise value.  Certainly, the
development of cavities may contribute to precipitate collapse
\cite{Khaleel,Taylor} but seems not to be the main cause. Maximum
elongation to fracture strongly depends on the strain rate, while the
evolution of cavities, when occurring, has been observed to be almost
independent of that variable \cite {Livesey1,Livesey2}. Moreover,
titanium superplastic alloys do not cavitate significantly
\cite{Cope2, Cope3}.

For decades, people have been interpreting their observations on the
basis of constitutive equations of the form $\sigma^n
=\dot\varepsilon^{\, m}/A$, relating the uniaxial applied stress
$\sigma$ with the superplastic strain rate $\dot\varepsilon$. The
coefficient $A$ depends on the microstructure and follows an Arrhenius
dependence with temperature. A variety of theories give this kind of
equation with constant $m$,
\cite{Ball,Mukherjee1,Langdon,Hayden,AshbyVerrall,Gifkins,Mukherjee2,Fukuyo}.
Nevertheless, experiments show that $\log\sigma$ and
$\log\dot\varepsilon$ are never linearly related, as implied by the
power law, but the curves display a sigmoidal shape, which is a
characteristic feature of the superplastic deformation, and the value
of the strain rate sensitivity $m$ that fits the data near the
inflexion point is usually a non-integral number which vary strongly
with strain rate and strain.  In many superplastic materials the large
ductility is observed even for $\log\dot\varepsilon$ far from the
inflexion point, where the slope $m$ has changed, and the strain rate
at the inflexion point shifts considerably with strain \cite{WangFu}.

Despite the success of some previous work in accurately predicting the
temperature dependent stress {\it vs.}~strain rate curves at zero or
small strain for a number of superplastic solids
\cite{Lagos0,Lagos1,Lagos2}, it rests to explain why are they
superplastic. Knowing the mechanisms that make the grains to slide
past each other and the successful prediction of plastic properties at
small strains are not enough to answer this question, and it is
necessary to investigate how grains flow over long paths when the
solid undergoes a finite deformation. The literature shows several
different theoretical attempts to explain superplastic phenomena, but
no one addresses the point of the modifications occurring inside the
material when strained up to fracture
\cite{Lagos0,Lagos1,Lagos2,Ball,Mukherjee1,Langdon,
Hayden,AshbyVerrall,Gifkins,Mukherjee2,Fukuyo}, and the dependence of
the stress {\it vs.} strain rate curves with strain.

We put forward here a general theoretical approach for the plastic
deformation of fine grained solids up to fracture strain. We assume
the grain boundary sliding mechanisms of Refs.~\onlinecite{Lagos1} and
\onlinecite{Lagos2}, but refine the study of the grain collective
motions in the spirit of Ref.~\onlinecite{Lagos3}. The latter is also
modified in order to interpret better the physics of the sliding
grains.  As a general conclusion, our results prove that the
deformation and fracture properties of superplastic solids are
governed by general laws amenable of analytical mathematical
description, simple and accurate enough to constitute a practical tool
for interpreting material tests. Though the assorted conglomerate of
possible structural defects may affect in specific situations, the
main mechanical behaviour of solids is determined by a few parameters
associated to the most basic structural properties.

\section{Stress tensor and plastic strain rates}
\label{strainrates}

\subsection{General considerations}
\label{genconsid}

At a scale much larger than the grain size, polycrystalline matter
lacks symmetry constrictions and periodicity, and displays same
average packing and properties in all directions, and over its whole
extention. However, despite this, assimilating a polycrystal to an
homogeneous and isotropic continuum may lead to gross errors, no
matter the scale, when dealing with it as a dynamical medium. It has
been proven before \cite{Lagos3}, and is revised in this section, that
the faceted nature of the structural constituents of a polycrystal,
the grains, influences the macroscopic plastic flow in a radical
manner.

To explain briefly how this does happen, assume that plastic flow
takes place by just grain sliding, and that grains readily accomodate
their shapes by slip, so the rate limiting process is the former.
Sliding of two adjacent grains occurs when the shear stress in the
plane of the common boundary exceeds a critical value $\tau_c$.
However, no matter how big the externally applied forces may be, shear
stresses vanish in planes whose normals are in the principal
directions of the stress tensor. By continuity, shear stresses in
planes whose normals are inside well--defined finite solid angles
around the principal directions are smaller than $\tau_c$. Hence, many
grains are impeded to slide simply because of the orientation of their
surfaces
\cite{Lagos3}. By this effect, the material medium ceases to be
dynamically isotropic, and cones are created by the very external
forces within which no displacement is allowed. These forbidden cones
destroy the kinematic balance and $\nabla\cdot\vec v\ne 0$, where
$\vec v$ represents the velocity field of the flowing medium. Hence,
mass density is not locally conserved in the overall plastic flow. The
consequent internal pressure build up helps deformation and, at a
critical strain, the material becomes mechanically unstable,
undergoing brittle or ductile fracture, depending on the precise
physical circumstances. The internal pressure $p$, which operates in
the local microscopic scale, keeps account of cumulated strain and
triggers fracture at a precise value of it.

\subsection{Relative velocities between adjacent grains and strain
rates} \label{grainkinematics}

The proper configuration variables in the macroscopic scale are the
components $\varepsilon_{ij}$ ($i,j=x,y,z$) of the strain tensor,
which, together with their time derivatives $\dot\varepsilon_{ij}$,
give a complete picture of the dynamical state of the system. However,
the evolution of $\varepsilon_{ij}$ in time is governed by forces
excerted between each pair of adjacent grains, and hence the dynamical
analysis demands to go first to the grain level. As the objective is
to return to the larger scale, the first task is to establish the
connection between the two descriptions.

Given the statistics of spatial orientations of the grain boundaries,
which may be assumed isotropic from the beginning, one can derive the
relation between $\dot\varepsilon_{ij}$ and the set of the relative
velocities $\Delta\vec v$ between pairs of adjacent grains. To show
this, temporarily place the origin O of the coordinate system at a
point of a grain surface. Then call $\vec v$ the velocity of a point P
in the positive $i$-coordinate axis, which is in another grain
boundary distant $x_i=nd$ from O, where $n$ is a large natural number
and $d$ the mean grain size. (In what follows it will be denoted
either $i,j=x,y,z$ or $i,j=1,2,3$, with $x_1\equiv x,\, x_2\equiv y,\,
x_3\equiv z $).

As there are $n$ contiguous grains between points O and P, $\vec v$ is
the sum

\begin{equation}
\vec v = \sum_{p=1}^n \Delta \vec v(p)+
\sum_{p=1}^n u_i(p)\hat e_i ,
\label{EC1}
\end{equation}

\noindent
where $p$ correlatively numbers the adjacent grains along the $i$-axis
and $\Delta \vec v(p)$ is the velocity of grain $p$ relative to grain
$p-1$ at the interface between them. The second term in the right hand
side of Eq.~(\ref{EC1}) accounts for the assumed deformability of the
grains, $\hat e_i$ is the unit vector along the $i$--axis and $u_i(p)$
stands for the expansion rate of grain $p$ along the $i$--direction.
However, we are considering plastic deformation processes in which
grains remain equiaxed in the mean and do not vary significantly their
average size. Therefore, on this basis, as $n$ is large we can set

\begin{equation}
\sum_{p=1}^n u_i(p)\hat e_i=0,
\label{EC1a}
\end{equation}

\noindent
which means that, statistically speaking, the grains neither stretch
nor shrink. Hence

\begin{equation}
\vec v = \sum_{p=1}^n \Delta \vec v(p)
\label{EC2}.
\end{equation}

If the solid is being deformed, the components of the displacement of
P in a time $\delta t$ are

\begin{equation}
v_i\delta t=\dot\varepsilon_{ii}x_i\delta t, \quad
v_j\delta t=\dot\varepsilon_{ji}x_i\delta t \quad (j\ne i).
\label{EC3}
\end{equation}

\noindent
Substituting $x_i=nd$ and Eq.~(\ref{EC2}) in Eqs.~(\ref{EC3}), it
is obtained that

\begin{equation}
\dot\varepsilon_{ji}=\frac{1}{nd}\sum_{p=1}^n \Delta v_j (p) ,
\label{EC4}
\end{equation}

\noindent 
where the $i$ dependence in the right hand side is implicit in the
direction along which the relative velocities are sampled. (Notice
that the present arguments leading to Eq.~(\ref{EC4}) do not assume
that the grains are rigid, as seems implicit in the demonstration
given in Ref.~\onlinecite{Lagos3}, but that they conserve mean size
and shape in the plastic flow).

If the material is homogeneous and isotropic in the macroscopic scale,
the $n$ plane grain interfaces intersecting OP are oriented in any
direction with the same probability. Therefore, one may interpret the
sum in Eq.~(\ref{EC4}) as an average which, as $n$ is large, runs over
all possible orientations of the grain boundaries intersecting OP.
However, vectors $\Delta\vec v$ represent relative velocities between
consecutive grains ordered along the positive $i$-coordinate axis.
Thus, to take this into account, the average must be performed over
all orientations in the hemispace $x_i\ge 0$. Equivalently, denoting
$\hat n$ the unit vector normal to the interface between two grains,
the sum in Eq.~(\ref{EC4}) can be replaced by an integral over all
directions $\hat n$ satisfying the condition $\hat n\cdot\hat e_i\ge
0$. Therefore, we can write

\begin{equation}
\dot\varepsilon_{ij}=\frac{1}{d}\langle\Delta v_j\rangle_i ,
\label{EC5}
\end{equation}

\noindent 
where symbol $\langle\dots\rangle_i$ means the average over all
orientations in the semispace $x_i\ge 0$. If the material is not
isotropic in the larger scale, for instance because the grains are not
equiaxed, the average (\ref{EC5}) should incorporate the proper weigth
factors.

\subsection{Relative velocities between adjacent grains and stresses}
\label{graindynamics}

\begin{figure}[h!]
\begin{center}
\includegraphics[width=5.5cm]{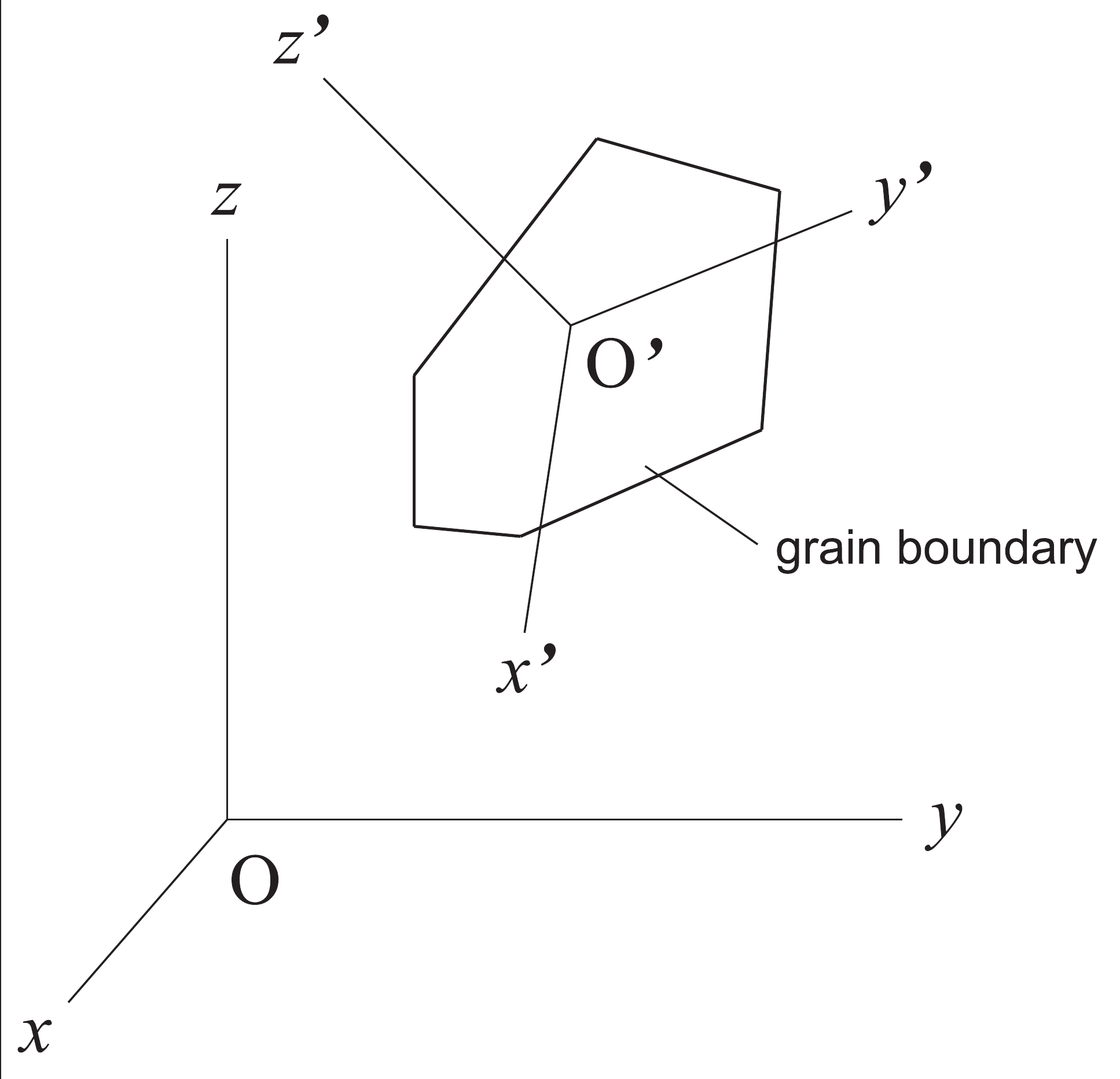}
\caption{\label{Fig1} Local reference system
$(x^\prime y^\prime z^\prime)$, with the $z^\prime$ axis normal to
the plane of the common boundary of two adjacent grains. The axes of
the $(xyz)$ frame of reference are in the principal directions of the
stress tensor.}
\end{center}
\end{figure}

The stress tensor

\begin{equation}
(\sigma_{ij}) = \left(
\begin{array}{ccc}
\sigma_x & 0 & 0 \\
0 & \sigma_y & 0 \\
0 & 0 & \sigma_z
\end{array}\right) , 
\label{EC6}
\end{equation}

\noindent
is diagonal in the main frame of reference $(xyz)$, with $\sigma_i$
($i=x,y,z$) being the principal stresses. To properly express the
physics at the grain scale, define a local frame of reference
$(x'y'z')$ whose $x'y'$ plane is in the common grain boundary plane of
two adjacent grains (Fig.~\ref{Fig1}). In the local reference system
the stress tensor reads

\begin{equation}
(\sigma_{i^\prime j^\prime})=R(\sigma_{ij})R^T ,
\label{EC7}
\end{equation}

\noindent
where $R$ is the rotation matrix that makes the transformation from
the $(xyz)$ to the $(x^\prime y^\prime z^\prime)$ frames, and $R^T$ is
the transposed of $R$.

As pointed out in Ref.~\onlinecite{Lagos1}, superplastic grain
boundary sliding seems closely related with elastic instabilities
occurring at interfaces in solids when subjected to overcritical
stresses \cite{Bellon, Srolovitz, Muller, Grinfeld, Williams, Genin,
Jones, Jesson, Searson, Hwang}. Monte Carlo simulation and analytical
treatment show that, in general, planar structures inside the bulk of
crystalline matter become rough when subjected to high enough
in--plane shear forces \cite{Bellon}. After the occurrence of the
elastic instability the locally induced normal stress field drives an
active atomic transport between the interface and the two adjacent
crystals. This bulk effect is closely related with the
Asaro--Tiller--Grinfeld instability, {\it i.~e.}~stress induced
roughening of solid surfaces \cite{Srolovitz, Muller, Grinfeld}, and
with the spontaneous roughening of thin films produced by strong
in--plane stresses arising from lattice mismatch with the substrate
\cite{Williams, Genin, Jones, Jesson, Searson, Hwang}.

It has been shown that the active atomic transport between the
buckled grain boundary and the two adjacent grains follows closed
paths, which makes the two grain surfaces to slide \cite{Lagos1}. The
relative velocity between the two sliding grains turns out to be
proportional to the in--plane components,
$\sigma_{x'z'}$ and $\sigma_{y'z'}$, of the shear stress operating in
the shared grain boundary plane, provided that the total in--plane
shear stress

\begin{equation}
\tau_{z'}=\sqrt{\sigma_{x'z'}^2+\sigma_{y'z'}^2}
\label{EC8}
\end{equation}

\noindent
be higher than the threshold stress $\tau_c$ for the elastic
instability of the grain boundary \cite{Lagos1, Lagos2, Lagos3}.
Hence

\begin{equation}
\begin{aligned}
&\Delta v_{i'}=
\begin{cases}
\mathcal{Q}\, \sigma_{i'z'}
\left( 1-\displaystyle\frac{\tau_c}{\tau_{(x'y')}}\right),
&\text{ if}\, \tau_{(x'y')}>\tau_c, \, i'=x', y',\\
0, &\text{ otherwise},  
\end{cases} \\
&\Delta v_{z'}\equiv 0,
\label{EC9}
\end{aligned}
\end{equation}

\noindent
where $\mathcal{Q}$ is a coefficient depending on the normal stresses
$\sigma_{i'i'}$ which will be discussed in detail in the next section.
The factor $(1-\tau_c/\tau_{z'})$ ensures that $\Delta\vec v =0$ for
$\tau_{z'}=\tau_c$.

\begin{figure}[h!]
\begin{center}
\includegraphics[width=6cm]{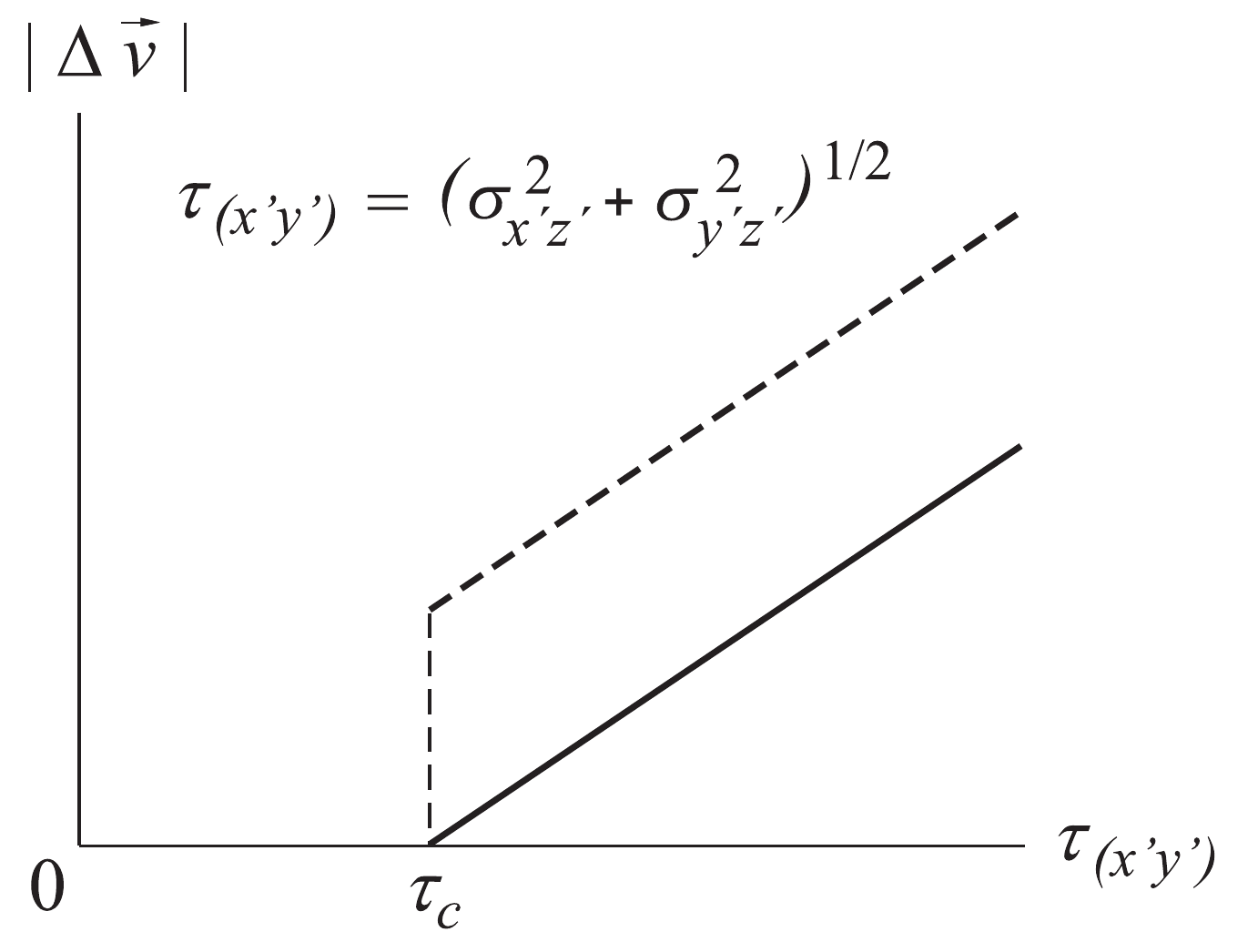}
\caption{\label{Fig2} The solid line represents the linear relation
assumed here between $|\Delta\vec v|$ and the total in--plane shear
stress. The dotted line represents the model of Ref.~\cite{Lagos3}.}
\end{center}
\end{figure}

The coefficient $\mathcal{Q}$ must not depend on the orientation of
the local frame of reference $(x^\prime y^\prime z^\prime)$, therefore
its dependence on the normal stresses is only via the invariant

\begin{equation}
p=-\frac{1}{3}(\sigma_{x'x'}+\sigma_{y'y'}+\sigma_{z'z'})=
-\frac{1}{3}(\sigma_x+\sigma_y+\sigma_z).
\label{EC9a}
\end{equation}

\noindent
This is also a necessary and sufficient condition for considering
$\mathcal{Q}(p)$ as independent of the shear stresses for any
orientation of the grain boundary. But for the critical condition
incorporated in Eqs.~(\ref{EC9}), $\mathcal{Q}$ may be seen as an
inverse grain boundary viscosity coefficient
\cite{AshbyVerrall, RajAshby, Wei}.

Notice that $\Delta v_{i'}$ in Eqs.~(\ref{EC9}) is related with an
irreversible shift between the two grains, and hence cannot describe
the variation of an elastic degree of freedom. In what follows,
elastic and plastic strains will be independent mechanical variables
(which nevertheless do interact). Grain elastic distortions are always
implicit in the subsequent developments, but plastic strains are the
dynamical variables because grains are the flowing entities. Elastic
strains of the grains will remain implicit in the stresses, which are
linearly related with elastic strains through the equations of the
theory of elasticity.

To put Eqs.~(\ref{EC9}) in a more practical way, we must perform the
matrix transformation (\ref{EC7}), choose the proper components of
$(\sigma_{i'j'})$ to insert them in Eqs.~(\ref{EC9}) and write
$\Delta\vec v$ in the local $(x'y'z')$ reference system, and then make
the inverse vector transformation to express $\Delta\vec v$ in the
main $(xyz)$ coordinate system. Hence, for $\tau_{z'}>\tau_c$ the
components $\Delta v_{i^\prime}$ are

\begin{equation}
\begin{aligned}
&\Delta v_{i^\prime}=
\mathcal{Q}\left[ R(\sigma_{ij})R^T\right]_{i^\prime z^\prime}
\left( 1-\displaystyle\frac{\tau_c}{\tau_{(x'y')}}\right),
\; i^\prime = x^\prime ,y^\prime ,\\
&\Delta v_{z^\prime} =0,
\label{EC10}
\end{aligned}
\end{equation}

\noindent 
where $\left[ \dots \right]_{i^\prime z^\prime}$ selects the $i^\prime
z^\prime$ component of the tensor. When expressed in the $(xyz)$
frame, the relative velocities (\ref{EC10}) are $(\Delta v_i)=R^T
(\Delta v_{i^\prime})$ or, more explicitly,

\begin{equation}
\Delta v_i=\sum_{i^\prime =x^\prime ,y^\prime}
R_{i^\prime i}\Delta v_{i^\prime} , \quad
i=x,y,z .
\label{EC11}
\end{equation}

\begin{figure}[h!]
\begin{center}
\includegraphics[width=5cm]{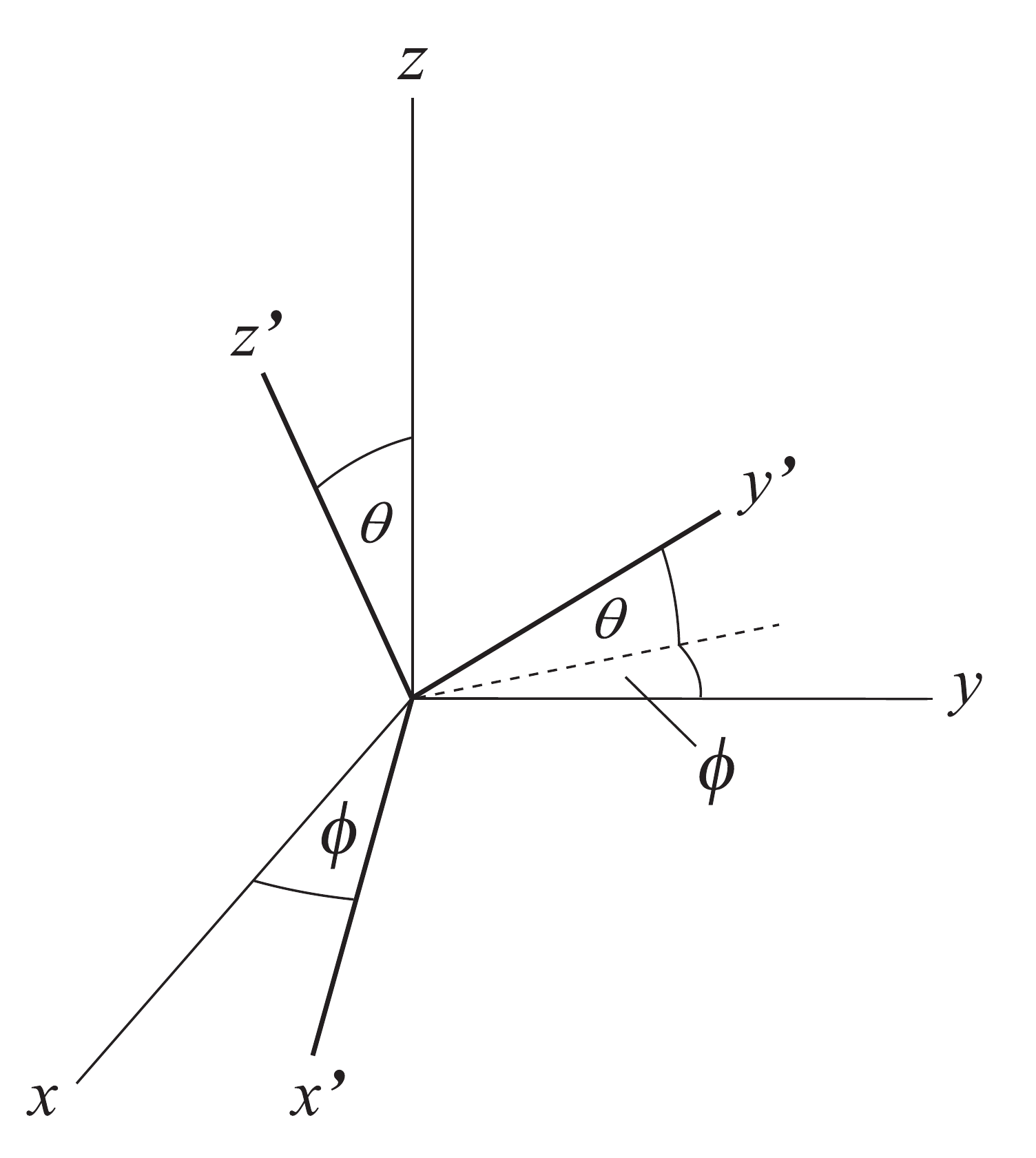}
\caption{\label{Fig3} Angles between the axes of the local
reference system $(x^\prime y^\prime z^\prime)$ and the
principal directions. The $x^\prime$ axis is in the $xy$
plane.}
\end{center}
\end{figure}

The local $(x^\prime y^\prime z^\prime)$ frame can be arbitrarily
rotated around the $z'$ axis without breaking the condition that $z'$
be normal to the grain boundary plane. Then, the $x^\prime$ axis can
always be chosen on the $xy$ plane, as in Fig.~\ref{Fig3}. Calling
$\phi$ the Euler angle between the $x^\prime$ and $x$ axes, and
$\theta$ the other Euler angle, going from the $z$ to the $z^\prime$
axis, the rotation matrix reads

\begin{equation}
R(\theta, \phi ) = \left( \begin{array}{ccc}
\cos\phi & \sin\phi & 0 \\
-\sin\phi \cos\theta & \cos\phi  \cos\theta & \sin\theta \\
\sin\phi \sin\theta & -\cos\phi \sin\theta & \cos\theta
\end{array}\right).
\label{EC12}
\end{equation}

Transforming $(\sigma_{ij})$ with the matrix (\ref{EC12}) one obtains,
in particular, that

\begin{equation}
\begin{aligned}
&\sigma_{x'y'}=(\sigma_x-\sigma_y)\sin\phi\cos\phi\sin\theta \\
&\sigma_{y'z'}=-(\sigma_x\sin^2\phi +\sigma_y\cos^2\phi
-\sigma_z)\sin\theta\cos\theta ,
\end{aligned}
\label{EC13}
\end{equation}

\noindent
and, making the inverse vector transformation (\ref{EC11}), the
components of the relative velocity between grains takes the explicit
form

\begin{equation}
\begin{aligned}
\Delta v_x &=\mathcal{Q} \sin\theta
\big[ (\sigma_x -\sigma_y)\sin\phi \cos^2\phi +(\sigma_x \sin^2 \phi\\
&+\sigma_y \cos^2\phi -\sigma_z)\sin\phi\cos^2\theta\big]
\left( 1-\frac{\tau_c}{\tau (\theta ,\phi)}\right)\\
\Delta v_y &=\mathcal{Q} \sin\theta
\big[ (\sigma_x -\sigma_y)\sin^2 \phi \cos\phi -(\sigma_x \sin^2 \phi\\
&+\sigma_y \cos^2 \phi -\sigma_z)\cos\phi\cos^2\theta\big]
\left( 1-\frac{\tau_c}{\tau (\theta ,\phi)}\right)\\
\Delta v_z &=-\mathcal{Q} (\sigma_x \sin^2 \phi\\
&+\sigma_y \cos^2 \phi -\sigma_z)\sin^2 \theta \cos\theta
\left( 1-\frac{\tau_c}{\tau (\theta ,\phi)}\right) .
\end{aligned}
\label{EC14}
\end{equation}

\noindent
Eqs.~(\ref{EC14}) hold when

\begin{equation}
\tau (\theta ,\phi)>\tau_c ,
\label{EC15}
\end{equation}

\noindent
with

\begin{equation}
\begin{aligned}
\tau &(\theta ,\phi) =
\big[ (\sigma_x-\sigma_y)^2\sin^2\phi\cos^2\phi\sin^2\theta \\
&+(\sigma_x\sin^2\phi +\sigma_y\cos^2\phi -\sigma_z)^2
\sin^2\theta\cos^2\theta\big]^{1/2} ,
\end{aligned}
\label{EC16}
\end{equation}

\noindent
otherwise $\Delta\vec v =0$.

\subsection{Strain rates and stresses}
\label{graindyn}

To simplify the equations, let us particularize to the important case
of uniaxial external stress $\sigma_z $. Far from the sample surfaces,
the material has cylindrical symmetry and $\sigma_x = \sigma_y $,
which will be assumed not to vanish for the sake of generality.
Eqs.~(\ref{EC14}) and (\ref{EC16}) greatly reduce and give

\begin{equation}
\begin{aligned}
\Delta\vec v &= \mathcal{Q}\, (\sigma_z - \sigma_x)
\sin\theta\cos\theta \\
&\times\bigg( 1-\frac{\tau_c}{|(\sigma_z
-\sigma_x)\sin\theta\cos\theta|}\bigg)
\,\hat v \quad (\sigma_x=\sigma_y),
\end{aligned}
\label{EC17}
\end{equation}

\noindent
where the unit vector $\hat v$ in the $(x,y,z)$ system is

\begin{equation}
\hat v=(-\sin\phi\cos\theta ,\cos\phi\cos\theta ,\sin\theta ).
\label{EC18}
\end{equation}

\begin{figure}[h!]
\begin{center}
\includegraphics[width=9cm]{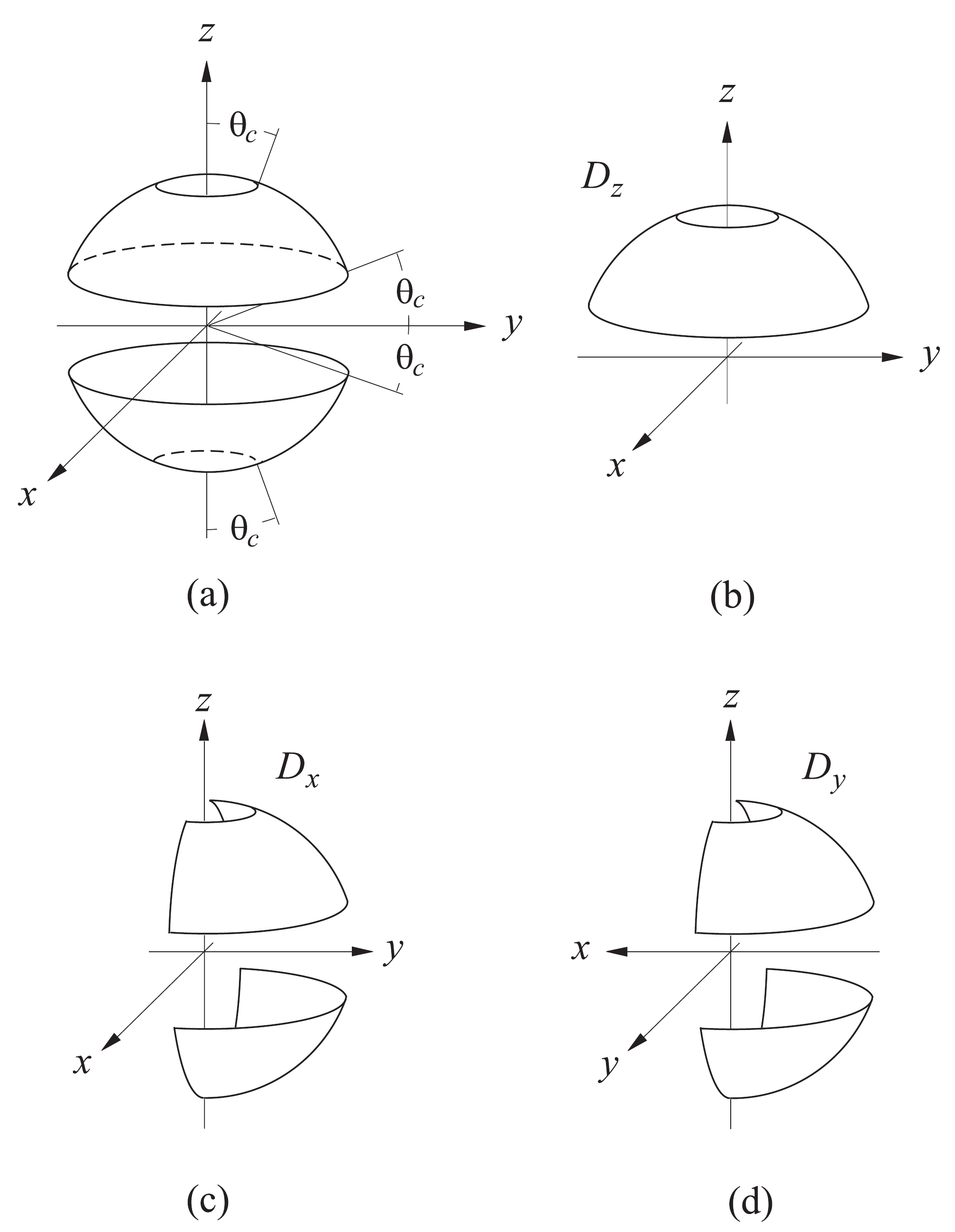}
\caption{\label{Fig4} (a) Grain boundaries whose normals are in the
spherical shells are the only ones allowed to slide, if the stress
field has cylindrical symmetry with respect to the $z$ axis. Grain
boundaries whose normals make angles $0\le \theta <\theta_c$,
$\pi/2-\theta_c <\theta <\pi/2+\theta_c$ and
$\pi-\theta_c <\theta <\pi$ with the $z$ axis do not contribute to
plastic flow. (b), (c) and (d) Integration domains for integral
(\ref{EC22}).}
\end{center}
\end{figure}

\noindent
The choice of the $x^\prime$ axis in the $xy$ plane proves to be
particularly convenient because, by the cylindrical symmetry, any
direction in the $xy$ plane is a principal direction, and then
$\sigma_{x^\prime z^\prime}=0$ for any $\theta$ and $\phi$. The
condition (\ref{EC15}) on the minimal shear stress $\tau_c$ becomes

\begin{equation}
|(\sigma_z - \sigma_x)\sin\theta\cos\theta| > \tau_c, \quad
(\sigma_x = \sigma_y),
\label{EC19} 
\end{equation}

\noindent
which determines a critical angle $\theta_c $, given by

\begin{equation}
\sin(2\theta_c) = \frac{2\tau_c}{|\sigma_z-\sigma_x|},
\label{EC20}
\end{equation}

\noindent
such that $\Delta\vec v \equiv 0$ for $|\sin 2\theta|<\sin
2\theta_c$. Therefore, under the restriction that $\sigma_x =
\sigma_y$, plastic flow is not allowed for $0\le \theta <\theta_c$,
$\pi/2-\theta_c <\theta <\pi/2+\theta_c$ and $\pi-\theta_c <\theta
<\pi$, as shown in Fig.~\ref{Fig4}(a).

To calculate the averages $\langle\Delta v_j\rangle_i$, appearing in
Eq.~(\ref{EC5}), we must integrate over all directions $(\theta ,
\phi)$ such that the unit vector $\hat n(\theta ,\phi)$, normal to
the grain boundary plane, has positive projection on the $x_i$
axis. Vector $\hat n$ is the unit vector $\hat k'$ along the $z'$ axis
of the local coordinate system, that is,

\begin{equation}
\hat n=(\sin\phi\sin\theta ,-\cos\phi\sin\theta ,\cos\theta ).
\label{EC21}
\end{equation}

\noindent 
In general,

\begin{equation}
\langle\Delta\vec v\rangle_i
=\frac{1}{2\pi}\int_{D_i}\, d\phi\, d\theta\,
\sin\theta\, \Delta\vec v(\theta ,\phi) ,
\label{EC22}
\end{equation}

\noindent
and the integration domains $D_i$ can be determined by examining the
geometry of the $(xyz)\rightarrow (x^\prime y^\prime z^\prime)$
transformation. They are (Fig.~\ref{Fig4}(b),(c) and (d))

\begin{equation}
\begin{aligned}
D_x : \;
&\theta\in [\theta_c,\pi/2-\theta_c]\cup[\pi/2+\theta_c,\pi-\theta_c],\\
&\phi\in [0,\pi],\\
D_y : \;
&\theta\in [\theta_c,\pi/2-\theta_c]\cup[\pi/2+\theta_c,\pi-\theta_c],\\
&\phi\in [\pi/2,3\pi/2],\\
D_z : \;
&\theta\in [\theta_c,\pi/2-\theta_c],\\
&\phi\in [0,2\pi].
\label{EC23}
\end{aligned}
\end{equation}

Integrating and combining with Eq.~(\ref{EC5}) one obtains

\begin{equation}
\dot\varepsilon_{zz} = 
\frac{\mathcal{Q}}{4d}\big[ (\sigma_z - \sigma_x) \cos(2\theta_c)
-\tau_c(\pi -4\theta_c)\big],
\label{EC24}
\end{equation}

\noindent
and

\begin{equation}
\begin{aligned}
\dot\varepsilon_{xx} = \dot\varepsilon_{yy} =
&-\frac{\mathcal{Q}}{8d}(\sigma_z-\sigma_x)
\bigg[ 1-\frac{4\theta_c}{\pi}
+\frac{\sin(4\theta_c)}{\pi} \bigg] \\
&+\frac{\mathcal{Q}\, \tau_c}{\pi d}\cos (2\theta_c).
\label{EC25}
\end{aligned}
\end{equation}

\noindent
Eqs.~(\ref{EC24}) and (\ref{EC25}) hold for $\sigma_z -
\sigma_x>0$.  When $\sigma_z - \sigma_x<0$ they must be modified
reversing the sign of the terms multiplied by $\tau_c$.

By Eq.~(\ref{EC20}), $\sigma_z - \sigma_x$ can be written in terms of
just $\theta_c$ and the constant $\tau_c$. Replacing in
Eqs.~(\ref{EC24}) and (\ref{EC25}), the strain rates turn out to be
entirely determined by $\theta_c$, which becomes a natural and
convenient auxiliary variable which varies in the range

\begin{equation}
0\le\theta_c\le\pi/4.
\label{EC26}
\end{equation}

\noindent
For finite $\tau_c$, $\theta_c=0$ corresponds to the asymptotic
situation in which the external forces applied on the sample are very
strong. For $\theta_c=\pi/4$, the forbidden directions extend over the
whole sphere and no grain is allowed to slide.

As the relative velocities, the strains $\varepsilon_{ij}$ are purely
plastic, but elastic strains are not disregarded and remain implicit
in the stresses, which are linearly related with elastic strains
through the equations of the theory of elasticity.

\section{Dilation rate and internal pressure build up}
\label{dilationrate}

From Eqs.~(\ref{EC24}) and (\ref{EC25}), the dilation rate
$\dot V/V=\dot\varepsilon_{xx}+\dot\varepsilon_{yy}
+\dot\varepsilon_{zz}$ turns out to be

\begin{equation}
\begin{aligned}
\frac{\dot V}{V} =
&-\frac{\mathcal{Q}}{4d}(\sigma_z - \sigma_x)
\bigg[ 1 -\cos(2\theta_c)-\frac{4\theta_c}{\pi} \\
&-\frac{\sin(4\theta_c)}{\pi}
+\bigg( \frac{\pi}{2}-2\theta_c\bigg)\sin (2\theta_c) \bigg]\, ,
\label{EC27}
\end{aligned}
\end{equation} 

\noindent
where $V$ stands for the volume of the sample.

Eq.~(\ref{EC27}) discloses the amazing feature that the dilation rate
does not vanish in general if the threshold stress for grain boundary
sliding $\tau_c\ne 0$. If $\tau_c=0$ Eqs.~(\ref{EC24}) and
(\ref{EC25}) reduce to

\begin{equation}
\begin{aligned}
&\dot\varepsilon_{zz} = 
\frac{\mathcal{Q}}{4d}(\sigma_z - \sigma_x), \\
&\dot\varepsilon_{xx} = \dot\varepsilon_{yy} =
-\frac{\mathcal{Q}}{8d}(\sigma_z - \sigma_x) \quad (\tau_c=0),
\label{EC28}
\end{aligned}
\end{equation}

\noindent
$\dot V/V=0$ and the material flows as a viscous liquid.

The dilation rate (\ref{EC27}) vanishes only in the extreme cases
$\theta_c=0$ and $\theta_c=\pi /4$, and for $\sigma_z - \sigma_x > 0$
is negative for any other value in between. This means that plastic
stretching involves a spontaneous compression of the grains, which can
only be elastic, and demands the ability of the material to support
some elastic volume reduction. As this ability is essentially limited,
plastic strain is limited as well, and a finite strain to failure is
implicit in Eq.~(\ref{EC27}). The single-crystal elastic
compressibility will determine how much the polycrystal could be
plastically elongated.

Another general consequence of Eq.~(\ref{EC27}) is that, in general,
the transversal stresses $\sigma_x= \sigma_y$ do not vanish in the
plastic deformation, even when the applied external force is strictly
uniaxial, and the material is isotropic in a scale much larger than
the grain size.  This may seem rather odd if one has in mind a static,
or quasistatic, picture of the deformation process. Such a picture is,
however, misleading. Plastic flow is an essentially dynamic process.

It must be noted that Eq.~(\ref{EC27}) deals with volume changes,
which are necessarily elastic, but has nothing to do with grain shape
variations. Our primary hypothesis is that grains are plastic, and
they readily accommodate their shapes to preserve matter continuity,
but grain boundary sliding is the rate limiting process. This is
consistent with accomodation by slip, since crystal deformation by
slip does not demand much additional stress once the critical resolved
shear stress (CRSS) has been exceeded. Then, if $\tau_c$ is greater
than the CRSS of the material then our hypothesis about grain
plasticity should hold. We are not very specific about the
accommodation mechanisms because grain plastic deformation is not the
rate limiting process, diffusional creep or dislocation creep may
contribute, but we assume that the main accomodation mechanism is by 
slip processes.

Settling $\sigma_x= \sigma_y=0$ for $\varepsilon_{zz}=0$ is a natural
initial condition. However, it will be shown that, as the plastic
stretching along the $z$ axis increases with time, the magnitude of
the transversal stresses $\sigma_x= \sigma_y$ monotonically increases
as well, with negative sign. Hence, the plastic elongation of the
sample produces a spontaneous compression in the transversal plane.
Neck formation prior to ductile fracture of polycrystalline bars can
be attributed to this circumstance.

\begin{figure}[h!]
\begin{center}
\includegraphics[width=6cm]{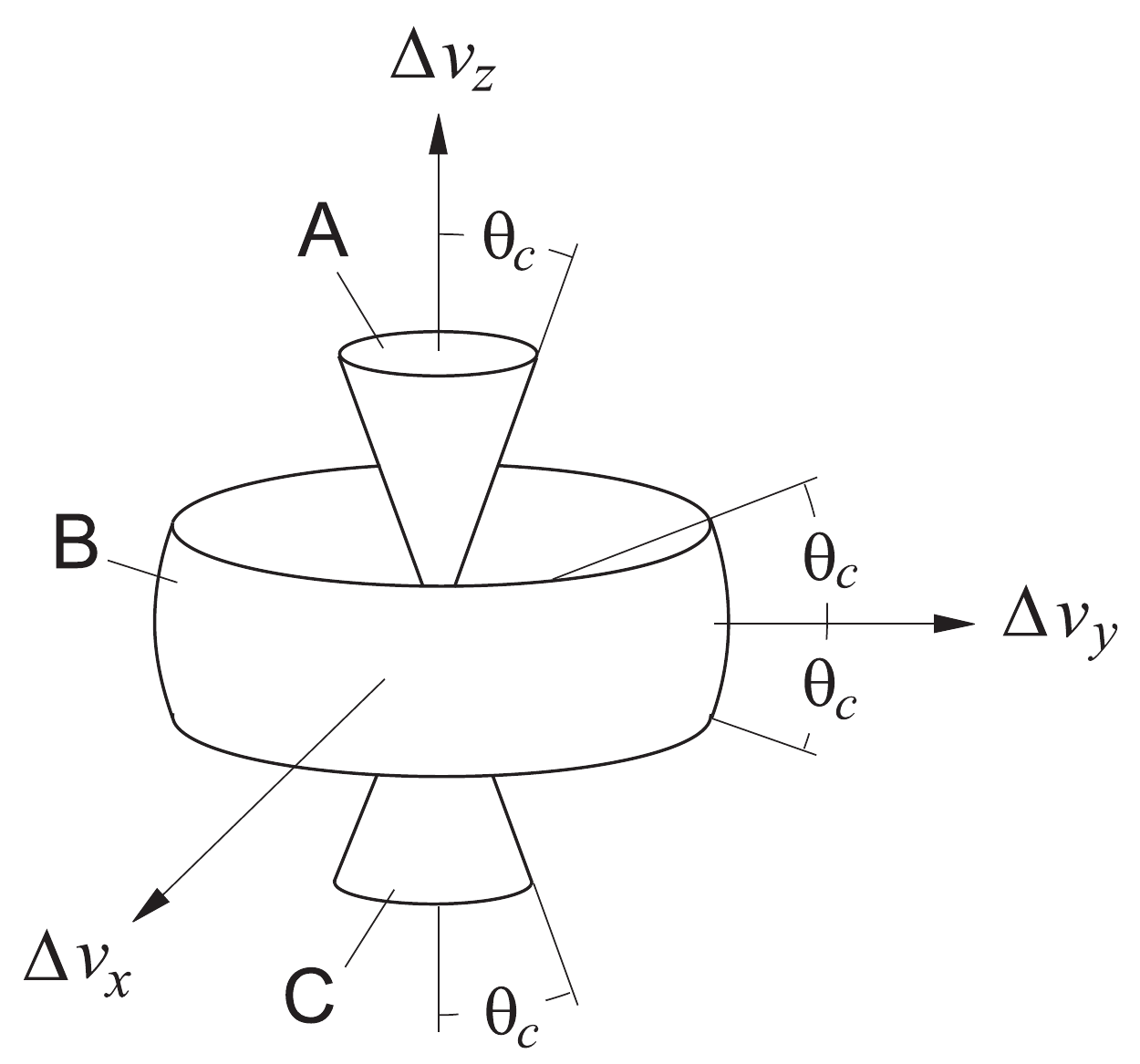}
\caption{\label{Fig5} The forbidden directions for $\Delta\vec v$.
Vectors $\Delta\vec v$ in the solid angles A and C correspond to grain
boundaries with $\hat n$ in the solid angle B, and vice--versa. Then,
there are more vectors supressed in A and C than in B.}
\end{center}
\end{figure}

The dilation rate (\ref{EC27}) can be understood in a qualitative
manner by a simple geometric argument. The relative speed $|\Delta\vec
v|$ between adjacent grains depends only on $\theta$ and is symmetric
with respect to $\theta =\pi /4$ in the top hemisphere, and with
respect to $\theta =3\pi /4$ in the bottom one
(Eq.~(\ref{EC17})). Fig.~\ref{Fig5} depicts the solid angles A, B and
C in which $\Delta\vec v\equiv 0$. Grain boundaries whose normals
$\hat n$ are in B correspond to banned relative velocities $\Delta\vec
v$ in A or C, and vice--versa. As the solid angle B is larger than the
sum of A and C, there are more supressed vectors $\Delta\vec v$ in A
and C than in B. Therefore, grain displacements in the $z$ direction,
or close to it, are less favoured than those occuring in directions
near the $xy$ plane. If no dilation takes place when $\theta_c = 0$
(i.~e.~$\tau_c =0$), some volume change should occur for $\tau_c\ne
0$. The same kind of argument serves to show that no dilation does
occur in two dimensions, and the area is conserved in the plastic
distortion of a strictly two--dimensional layer.

We will examine the case of the uniaxial plastic strain of a bulk
sample along the $z$ axis at a strain rate
$\dot\varepsilon\equiv\dot\varepsilon_{zz}$. The only externally
applied stress is $\sigma\equiv\sigma_z$. We will denote also
$p=-(\sigma_x+\sigma_y+\sigma_z)/3$ for the pressure. The transversal
stresses $\sigma_x=\sigma_y$ are assumed finite to give a chance for
internal stresses to develop. Of course, cylindrical symmetry demands
that the point into consideration be far enough from the surfaces.
With this notation

\begin{equation}
\sigma_x =\sigma_y =-\frac{1}{2}(\sigma +3p),
\quad \sigma_z -\sigma_x =\frac{3}{2}(\sigma +p).
\label{EC28a}
\end{equation}

Recalling Hooke's law 

\begin{equation}
\frac{\Delta V}{V}=-\frac{1}{B}\, p,
\label{EC29}
\end{equation}

\noindent
where $B$ is the bulk elastic modulus, the dilation rate
reads

\begin{equation}
\frac{\dot V}{V}=-\frac{1}{B}\frac{dp}{dt}.
\label{EC30}
\end{equation}

\noindent
Replacing the identity
$\dot\varepsilon\, dp/d\varepsilon=dp/dt$ and
Eq.~(\ref{EC27}) in Eq.~(\ref{EC30}),

\begin{equation}
\begin{aligned}
\frac{dp}{d\varepsilon}=
&s\mathcal{Q}\, \frac{B\tau_c}{2d\dot\varepsilon}
\bigg[\frac{1-\cos (2\theta_c)}{\sin (2\theta_c)}\\
&-2\theta_c\bigg( 1+\frac{2}{\pi\sin (2\theta_c)}\bigg)
-\frac{2\cos (2\theta_c)}{\pi}+\frac{\pi}{2}\bigg],
\label{EC31}
\end{aligned}
\end{equation}

\noindent
where use was made of Eq.~(\ref{EC20}) to eliminate
$\sigma_z-\sigma_x$ and put the right hand side in terms of just
$\theta_c$. The factor $s$ is

\begin{equation}
s=\left\{ \begin{array}{ll}
1, &\text{if $\sigma_z-\sigma_x>0$} \\
-1, &\text{if $\sigma_z-\sigma_x<0$} .
\end{array} \right.
\label{EC32}
\end{equation}

\noindent
For traction ($s=1$), the right hand side of Eq.~(\ref{EC31}) is
positive for $0\le\theta_c\le\pi/4$, and hence the pressure $p$
increases monotonically with strain.

Elastic degrees of freedom, implicit in $\dot V/V$, are brought to the
fore once more by Eq.~(\ref{EC27}). Eq.~(\ref{EC30}) expresses the
volume variation (the trace of the elastic strain rate tensor) in
terms of the pressure $p$ (essentially the trace of the stress
tensor). Hence it allows to return to the scheme in which the elastic
distortions are kept implicit in the stresses. The variation of $p$ in
the plastic deformation indicates that elastic and plastic degrees of
freedom are coupled.

Written in terms of just the critical angle $\theta_c$,
Eq.~(\ref{EC24}) for the strain rate becomes

\begin{equation}
\dot\varepsilon =s\,\mathcal{Q}\frac{\tau_c}{2d}\left[ \cot (2\theta_c)
+2\theta_c-\frac{\pi}{2}\right],
\label{EC33}
\end{equation}

\noindent
with

\begin{equation}
\theta_c=\frac{1}{2}\arcsin\left(\frac{4\tau_c}{3|\sigma +p|}
\right).
\label{EC33a}
\end{equation}

\noindent
In what follows we will consider just traction, and $s$ will be
omitted.

At the start of the deformation process $\varepsilon =0$ and
$\sigma_x=\sigma_y=0$. Then the initial condition written in the new
notation is

\begin{equation}
p=-\frac{\sigma}{3} \quad\text{for} \quad\varepsilon =0.
\label{EC34}
\end{equation}

\section{The coefficient $\mathcal{Q}$} \label{slidingcoeff}

The coefficient $\mathcal{Q}$, governing grain boundary sliding, was
studied in detail in Refs.~\cite{Lagos1} and \cite{Lagos2}. It was
shown there that the stress dependence of $\mathcal{Q}$ is only
through the pressure $p$, and varies with temperature and grain size.
It obeys the rather simple law

\begin{equation}
\frac{\mathcal{Q}}{4d}=C_0\frac{\Omega^*}{k_BT}
\exp\left( -\frac{\epsilon_0+\Omega^*p}{k_BT}\right)
\label{EC35},
\end{equation}
 
\noindent
where $k_B$ is the Boltzmann constant, $T$ the temperature, the
coefficient $C_0$ depends only on the grain size $d$, the constant
$\epsilon_0$ is the energy necessary for evaporating a crystal vacancy
from the grain boundary, and $\Omega^*$ measures the sensitivity of
this energy to stress.

\begin{figure}[h!]
\begin{center}
\includegraphics[width=7.5cm]{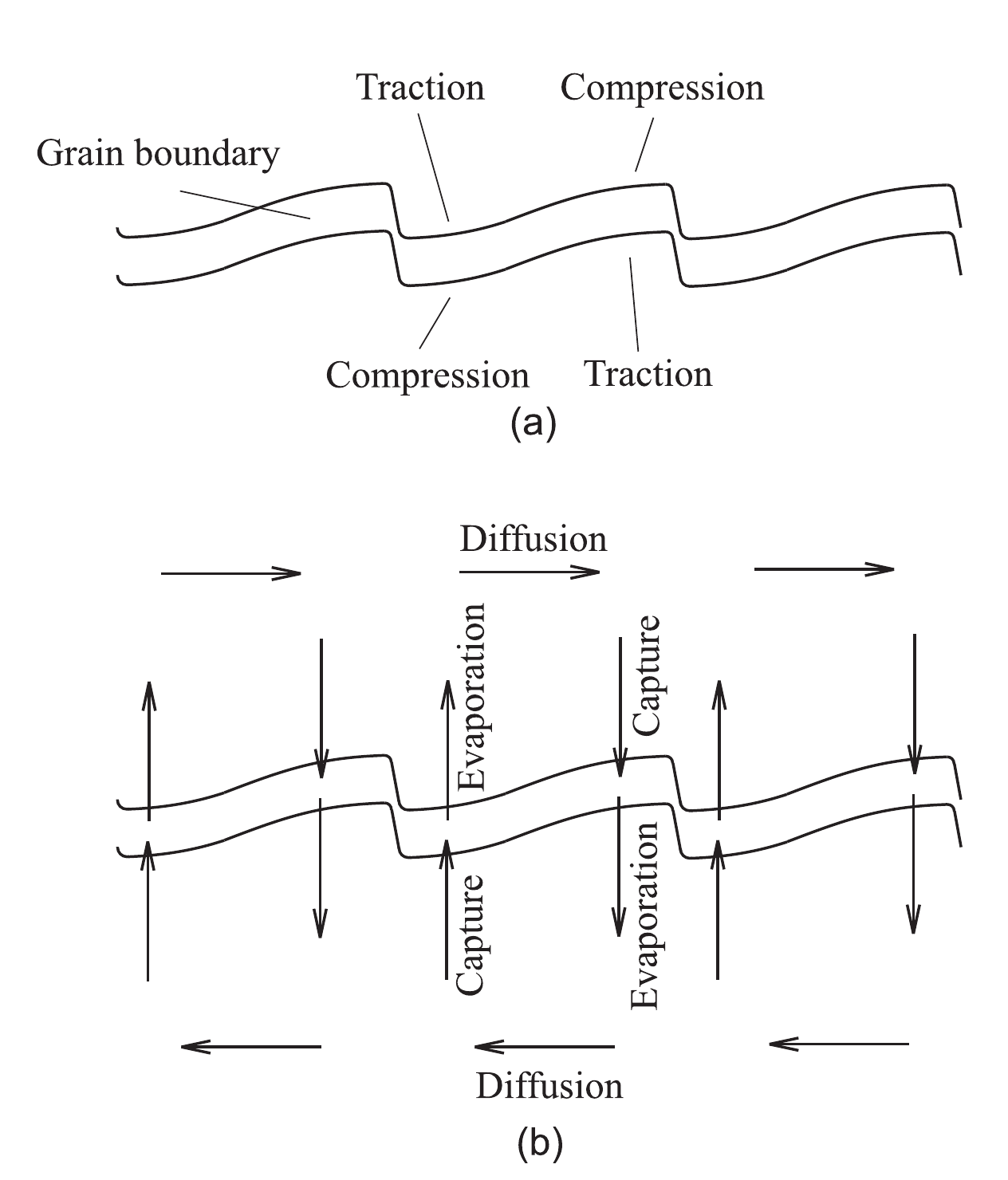}
\caption{\label{Fig6} (a) Schematic representation of a buckled grain
boundary, and the induced periodic stress field alternating
compression and traction regions in the adjacent crystals. (b) The
grain boundary releases and captures crystal vacancies in the
tractioned and compressed regions, respectively, producing this way
closed paths of the point defects. The diffusion of vacancies,
equivalent to an atomic counterflow, have opposite senses in the two
grains.}
\end{center}
\end{figure}

The derivation of expression (\ref{EC35}) has been already published
\cite{Lagos1,Lagos2}, but a brief explanation of the physics involved
is in order here. The mechanical analysis of a stressed polycrystaline
material must distinguish intergranular and crystalline matter because
they have different mechanical properties \cite{Lagos1,Lagos2}. A
shear stress greater than a critical value applied in the plane of a
grain boundary should buckle it, producing either a boundary
corrugation or the formation of a periodic series of trenches. This
localized deformation induces a periodic normal stress field that
alternates compression and traction on the adjacent crystal surfaces,
as shown schematically in Fig.~\ref{Fig6}(a). Grain boundaries are
efficient sinks and sources for vacancies and the periodic stress
field induced by the buckled boundary yields a periodic variation of
the equilibrium value for the concentration of crystal vacancies. The
grain boundary then evaporates and condenses the point defects in
alternate sectors, producing streams of the defects in closed loops
that cross the boundary and involve the two adjacent crystals
(Fig.~\ref{Fig6}(b)). These closed loops perform as the driving
pulleys of a conveyor belt and the phenomenon provides either an
accommodation mechanism and the driving force for crystal sliding
\cite{Lagos1,Lagos2}.

As explained, the grain boundary elastic instability causes a strongly
fluctuating stress field, which in the macroscopic scale averages to
zero. Be $\epsilon_B$ the energy necessary for the grain boundary to
release a vacancy into one of the two crystals adjacent to it, when
the system is unstrained. When the grains are strongly stressed, this
energy is expected to change linearly with stress, as long as
$\epsilon_B$ is much larger than its expected variations. Denoting
$\sigma_{ij}^1$ the characteristic amplitude of the induced stress
field spatial fluctuations, the relative concentration of defects in
two neighbouring regions of maximal and minimal stress is proportional
to

\begin{equation}
\exp \left[ -\frac{\epsilon_B+\Omega^*(p+p^1)}{k_BT}\right]
-\exp \left[ -\frac{\epsilon_B+\Omega^*(p-p^1)}{k_BT}\right],
\label{EC36}
\end{equation}
 
\noindent
where $p^1=-(\sigma_{xx}^1+\sigma_{yy}^1+\sigma_{zz}^1)/3$ is the
contribution from the induced stress field to the pressure. Expression
(\ref{EC36}) depends only on the normal stresses $\sigma_{xx}$,
$\sigma_{yy}$ and $\sigma_{zz}$ because it is expected that the energy
of a vacancy will not be significantly affected by the shear stresses,
which do not change the volume of the hole, but only its shape.

The flow of defects between these two regions is proportional to the
spatial concentration variation (\ref{EC36}), which can be written as

\begin{equation}
D\frac{\Omega^*p^1}{k_BT}
\exp\left( -\frac{\epsilon_B+\Omega^*p}{k_BT}\right)
\label{EC37}
\end{equation}
 
\noindent
for $p^1\ll p$. In Eq.~(\ref{EC37}) $D=D_0\exp[-\epsilon_a/(k_BT)]$ is
the diffusion coefficient for crystal vacancies, with $\epsilon_a$
being the corresponding activation energy. No matter the details of
the process, one can expect that the speed of the boundary sliding be
proportional to the rate at which the assisting vacancies flow along
the crystal boundaries, and expression (\ref{EC37}) leads to the main
Eq.~(\ref{EC35}) with $\epsilon_0 =\epsilon_B+\epsilon_a$.

By the time, direct experimental evidence was supplied by Vetrano {\it
et al.} \cite{Vetrano1, Vetrano2} for the deformation induced
supersaturation of vacancies in the vicinity of many grain boundaries
of an Al--Mg--Mn alloy during superplastic deformation. Rapid
quenching of the sample while being superplastically deformed revealed
the formation of very unstable nano--cavities in a great fraction of
the grain boundaries, which rapidly coalesce and disappear on moderate
heating. The authors show they start forming after cooling and
subsequent mounting of the sample in the transmission electron
microscope, and conclude that the cavities provides evidence of a
supersaturation of crystal vacancies near the sliding grain
boundaries.  The defects condense into voids when sliding abruptly
ceases, and the just released vacancies are not recaptured by the
grain boundaries.  Hence grain boundary sliding is associated to the
observed vacancy excess, which provides evidence of the capital role
of these defects in the deformation process.

Some aspects of the mechanism for grain boundary sliding introduced in
Refs.~\onlinecite{Lagos1} and \onlinecite{Lagos2}, and reviewed above
in a more compact approach, resembles the one put forward earlier by
Raj and Ashby \cite{RajAshby}. In both the shared grain boundary is
taken as a material medium able to release and capture crystal
vacancies, to or from the two grains, causing a relative motion of
them. Also, in both theoretical approaches the in--plane shear force
induces a strongly varying normal stress field localized in the
boundary region and, as a consequence, the less ordered intergrain
matter evaporates and condenses vacancies in a periodic series of
sources and sinks. The main difference is in the physical nature of
the sources and sinks for the point defects. In
Refs.~\onlinecite{Lagos1} and
\onlinecite{Lagos2} the sources and sinks are originated by the
buckling and trenching of the initially plane grain boundary (with
eventual steps) by effect of the overcritical shear stress applied to
it. Raj and Ashby \cite{RajAshby} assume an uneven grain boundary
subjected to a shear force. Sources and sinks are caused by the
variation of the normal stresses induced in the opposite sides of the
permanent grain boundary irregularities.

\section{The constitutive equation for small strain}
\label{constitutive}

\subsection{The equation} \label{actualequation}

When the plastic distortion just starts, $\varepsilon\approx 0$ and
holds the initial condition (\ref{EC34}), $p=-\sigma/3$. Then

\begin{equation}
\theta_c=\frac{1}{2}\arcsin\left( \frac{2\tau_c}{|\sigma|}\right)
\label{EC38}
\end{equation}

\noindent
and, replacing this and Eq.~(\ref{EC35}) in Eq.~(\ref{EC33}),

\begin{equation}
\begin{aligned}
\dot\varepsilon =
&2C_0\frac{\Omega^*\tau_c}{k_BT}
\left[\sqrt{\left(\frac{\sigma}{2\tau_c}\right)^2-1}
+\arcsin\left( \frac{2\tau_c}{\sigma}\right)-\frac{\pi}{2}\right]\\
&\times\exp\left( -\frac{\epsilon_0-\Omega^*\sigma /3}{k_BT}\right)
\qquad (\varepsilon\approx 0).
\label{EC39}
\end{aligned}
\end{equation}

Eq.~(\ref{EC39}) gives the plastic strain rate $\dot\varepsilon$ of an
axially symmetric sample at temperature $T$ under the action of the
uniaxial tensile stress $\sigma$. Hence it is the constitutive
equation of the material for the established conditions.

The discussion on grain sliding of sections \ref{strainrates} and
\ref{dilationrate} is valid for any polycrystalline solid. Also, there
is no reason to think that the grain boundary sliding mechanisms
leading to Eq.~(\ref{EC35}) for the coefficient $\mathcal{Q}$ be
exclusive to superplastic grain sliding. Hence, in principle,
Eq.~(\ref{EC39}) is expected to hold for the deformation of any fine
grained polycrystalline solid, fine enough to ensure that grain
boundary sliding be the rate limiting deformation process.

However, the large amount of experimental data available in the
literature on the superplastic materials makes them specially suited
for testing Eq.~(\ref{EC39}). The data for very small strain rates, in
the range $10^{-5}$ to $10^{-3}\,\text{s}^{-1}$, are specially
valuable because the effect of the critical stress $\tau_c$ on the
stress--strain rate curves is particularly notorius in it. Such small
strain rates are not usual in tensile tests of normal, not
superplastic, materials.

\subsection{Comparison with previous results and with experiment}
\label{previousequations}

To appreciate how $\tau_c$ affects the strain {\it vs.}~strain rate
curves, consider Eq.~(\ref{EC39}) in the asymptotic limit of very
large tensile stress $\sigma$, such that $\tau_c/\sigma\rightarrow
0$. The equation reduces to

\begin{equation}
\dot\varepsilon =
C_0\frac{\Omega^*\sigma}{k_BT}
\exp\left( -\frac{\epsilon_0-\Omega^*\sigma /3}{k_BT}\right)
\quad (\varepsilon\approx 0, \, \frac{\tau_c}{\sigma}\approx 0),
\label{EC40}
\end{equation}

\noindent
which is the result given in Ref.~\onlinecite{Lagos1}, in which the
effect of the critical stress $\tau_c$ discussed in sections
\ref{strainrates} and \ref{dilationrate} was passed over. This
produces no significant consequence when comparing the theoretical
results with data on alloys having small $\tau_c$, as is the case of
the aluminium Al--7475 samples whose mechanical testing was reported
by Hamilton \cite{Hamilton1}. However, the data on the same alloy
Al--7475 prepared by Pilling and Ridley \cite{Pilling1}, aluminium
Al--8090 SPF \cite{Pilling1} or titanium Ti--6Al--4V
\cite{Cope1, Cope2, Cope3}, do show some disagreement with
Eq.~(\ref{EC40}) for strain rates below $10^{-4}\, \text{s}^{-1}$.

\begin{figure}[h!]
\begin{center}
\includegraphics[width=8cm]{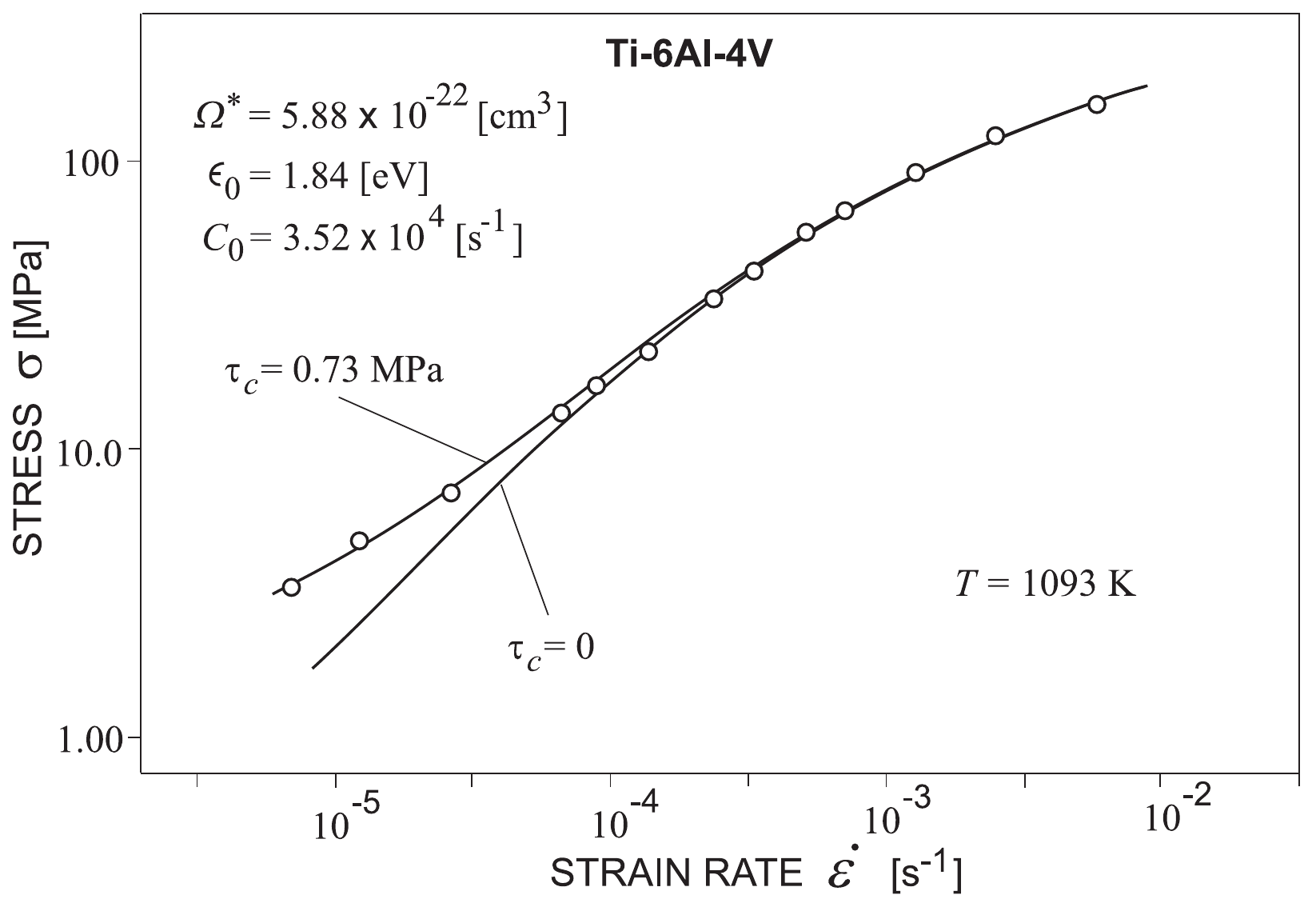}
\caption{\label{Fig7} Curves given by Eqs.~(\ref{EC39}) and
(\ref{EC40}), showing the effect of the critical shear stress
$\tau_c$. Values for the parameters are shown in the insets. Circles
represent data of Cope {\it et al.}, Refs.~\onlinecite{Cope2, Cope3,
Cope1}, on the superplastic titanium alloy Ti--6Al--4V, strained at
constant true strain rates and $T=1093\,\text{K}$.}
\end{center}
\end{figure}

As an illustrative example, Fig.~\ref{Fig7} shows the excellent fit
given by Eq.~(\ref{EC39}) to the data of Cope {\it et al.} on
Ti--6Al--4V \cite{Cope2, Cope3, Cope1} together with the curve
representing Eq.~(\ref{EC40}) with the same values for the parameters.
The latter has negative concavity over the whole range of
$\dot\varepsilon$ and does not exhibit the characteristic inflexion
point, which is traditionally taken as a distinctive feature of
superplasticity. However, both curves become practically the same for
$\dot\varepsilon >5\times 10^{-4}\,\text{s}^{-1}$.

Figure \ref{Fig7} also illustrates the usefulness of very low strain
rate data. As will be seen in the next sections, the critical shear
stress $\tau_c$ has important consequences in the plastic deformation
at finite strains and high or low strain rates, as fracture, for
example, but does not produce any easily recognizable feature in the
test curves. On the contrary, at low enough $\dot\varepsilon$, tensile
stresses are close to $2\tau_c$, and the solid angles A, B and C in
Fig.~\ref{Fig5} have significant magnitudes, producing the inflexion
point shown in Fig.~\ref{Fig7}. As the deformation mechanisms
considered here are in principle not exclusive to superplastic
materials, we guess that the strain {\it vs.}~strain rate curves for
normal ductile materials should exhibit and inflexion point as well,
at low enough strain rates.
 
In a subsequent paper \cite{Lagos2}, Eq.~(\ref{EC40}) was modified in
the way

\begin{equation}
\dot\varepsilon =C_0\frac{\Omega^*(\sigma -\sigma_0)}{k_BT}
\exp\left( -\frac{\epsilon_0-\Omega^*\sigma/3}{k_BT}\right),
\label{EC41}
\end{equation}  

\noindent
and an excellent agreement with the experimental results was achieved
at low, intermediate and high $\dot\varepsilon$. However, the
introduction of the threshold tensile stress $\sigma_0$ was based on
just intuitive physical arguments, and justified only by the good
results it produced.

\begin{figure}[h!]
\begin{center}
\includegraphics[width=6.5cm]{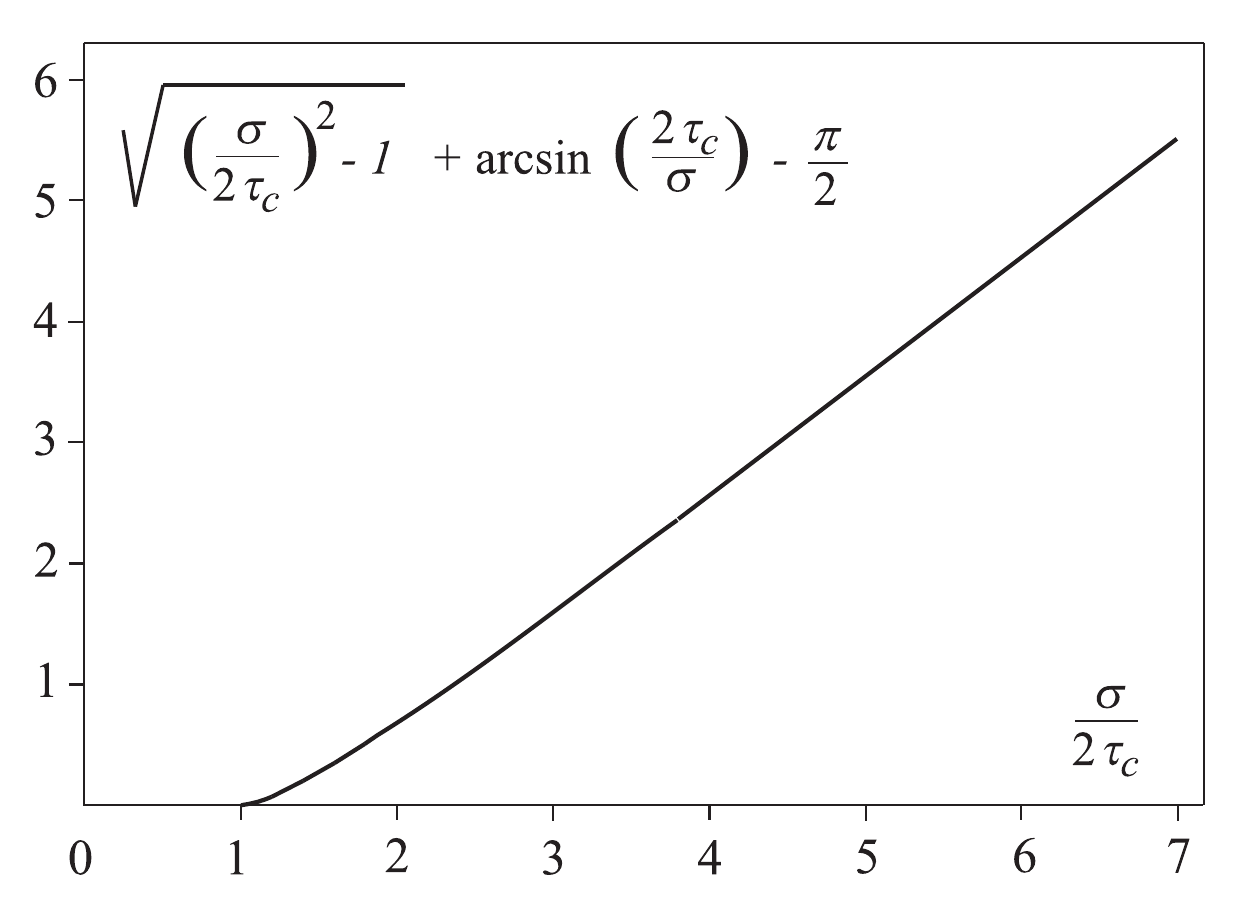}
\caption{\label{Fig8} The pre--exponential factor in Eq.~(\ref{EC39})
behaves like $\sigma -\sigma_0$ with $\sigma_0 =2\tau_c$.}
\end{center}
\end{figure}

Fig.~\ref{Fig8} explains the success of Eq.~(\ref{EC41}). It shows
that, despite its more involved appearance, the pre--exponential
factor in Eq.~(\ref{EC39}) is almost linear in $\sigma$, and vanishes
for $\sigma =2\tau_c$, which plays the role of $\sigma_0$ in
Eq.~(\ref{EC41}). Therefore, Eq.~(\ref{EC41}) constitutes a very good
approximation of the exact Eq.~(\ref{EC39}), and the same accurate fit
of the experimental data on Al--7475, Al--8090 SPF \cite{Pilling1},
and Ti--6Al--4V \cite{Cope2, Cope3, Cope1} attained with the former
\cite{Lagos2} is to be expected for the latter, but now with no
empirically justified modification.

\begin{figure}[h!]
\begin{center}
\includegraphics[width=8cm]{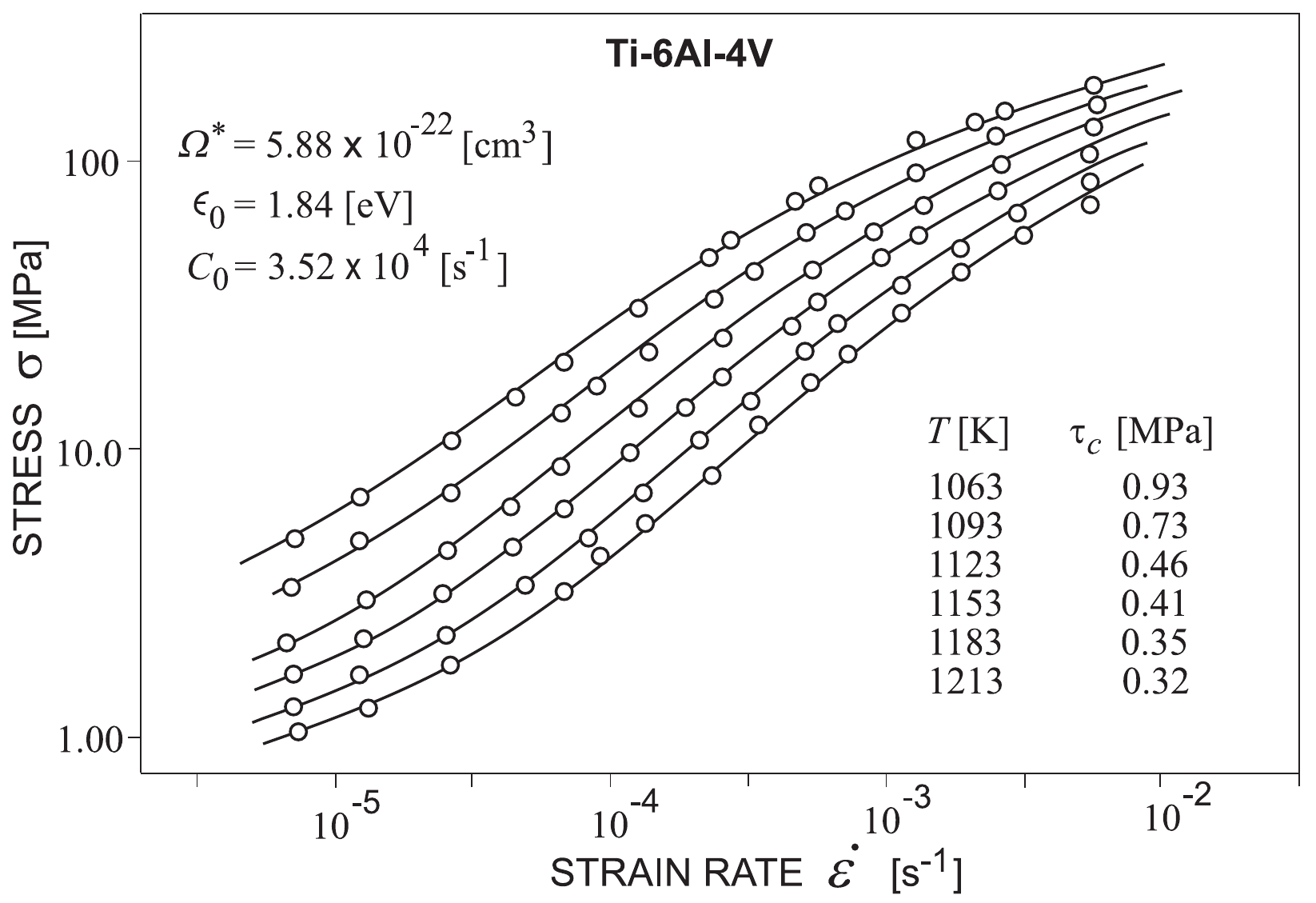}
\caption{\label{Fig9} Stress--strain rate curves for Ti--6Al--4V at
six temperatures. Circles represent data of Cope {\it et al.},
Refs.~\onlinecite{Cope2, Cope3, Cope1}, and solid lines depict
Eq.~(\ref{EC39}) for the parameters shown in the inset. The parameter
$\tau_c$ exhibits some temperature dependence.}
\end{center}
\end{figure}

Fig.~\ref{Fig9} shows the data of Cope {\it et al.}
\cite{Cope2, Cope3, Cope1} on the superplastic titanium alloy
Ti--6Al--4V at six temperatures (circles).  The solid lines represent
Eq.~(\ref{EC39}). The constants $\Omega^*$, $\epsilon_0$ and $C_0$
have the same values for the six curves, and are shown in one of the
insets. Notice that the values for $\Omega^*$, $\epsilon_0$ and $C_0$
used here in relation to Eq.~(\ref{EC39}) are the same than those used
in Ref.~\onlinecite{Lagos2} to insert in Eq.~(\ref{EC41}), with the
same success. The critical shear stress $\tau_c$ evidences some
dependence with temperature, as shown in the other inset.

The critical shear stress $\tau_c$ is expected to be temperature
dependent. It may be equal to either the smaller of the two critical
stresses associated to the grain boundary elastic instabilities
studied in Refs.~\onlinecite{Lagos1} and \onlinecite{Lagos2}, or the
critical resolved shear stress (CRSS) of the material when it is
greater than the critical stresses for the elastic instabilities. The
critical stresses for the elastic instabilities depend on the elastic
modulus of the material, which varies with temperature, and the CRSS
is temperature dependent as well when the temperature is high enough.
For Ti-6Al-4V it seems that $\tau_c$ is the CRSS. 

\section{Finite deformation at constant strain rate}
\label{finitestrain}

\subsection{The strain function} \label{strainfunction}

Deformation up to a finite strain $\varepsilon$ can be accomplished in
an indefinite number of inequivalent ways because of the other two
variables involved in the process, $\sigma$ and $\dot\varepsilon$.
Establishing the time dependence of one of them, or its dependence on
the intermediate values assumed by $\varepsilon$, would be enough to
determine the process. The simplest conditions are to keep
$\dot\varepsilon =\,\text{constant}$, as is usually done when testing
superplastic materials, or set $\sigma =\,\text{constant}$, which is
the common practice for normal ductile materials. In this section we
will assume the former.

Eq.~(\ref{EC31}) can be put in differential form as

\begin{equation}
\begin{aligned}
d\varepsilon =&\frac{2d\dot\varepsilon}{B\tau_c}
\bigg(\frac{1-\cos (2\theta_c)}{\sin (2\theta_c)}-2\theta_c\\
&-\frac{4\theta_c}{\pi\sin (2\theta_c)}
-\frac{2\cos (2\theta_c)}{\pi}+\frac{\pi}{2}\bigg)^{-1}
\frac{dp}{\mathcal{Q}(p)},
\label{EC42}
\end{aligned}
\end{equation}

\noindent
where $\dot\varepsilon$ is considered as a given constant. On the
other hand, writting Eq.~(\ref{EC33}) in the form

\begin{equation}
\frac{2d\dot\varepsilon}{\tau_c}\frac{1}{\mathcal{Q}}=
\cot (2\theta_c)+2\theta_c-\frac{\pi}{2}
\label{EC43}
\end{equation}

\noindent
and, derivating with respect to $p$ with $\dot\varepsilon$ constant,
one has that

\begin{equation}
\frac{2d\dot\varepsilon}{\tau_c}\frac{1}
{\mathcal{Q}^2}\frac{d\mathcal{Q}}{dp}=
\cot^2(2\theta_c)\frac{d\theta_c}{dp} .
\label{EC44}
\end{equation}

\noindent
Derivating also Eq.~(\ref{EC35}), the alternate expression for
$d\mathcal{Q}/dp$

\begin{equation}
\frac{d\mathcal{Q}}{dp}=-\frac{\Omega^*}{k_BT}\mathcal{Q}
\label{EC45}
\end{equation}

\noindent
is inferred. Replacing it in Eq.~(\ref{EC44}) and writting the result
in differential form gives

\begin{equation}
\frac{dp}{\mathcal{Q}}=
-\frac{\tau_ck_BT}{d\Omega^*\dot\varepsilon}
\cot^2(2\theta_c)\, d\theta_c .
\label{EC46}
\end{equation}

\noindent
Combining Eqs.~(\ref{EC42}) and (\ref{EC46}), and integrating, one
finally obtains

\begin{equation}
\begin{aligned}
\varepsilon =
&-\frac{2k_BT}{B\Omega^*}\int_{\theta_0}^{\theta_c}d\theta\,
\cot^2(2\theta)\bigg[\frac{1-\cos (2\theta)}{\sin (2\theta)}
-2\theta\\
&-\frac{4\theta}{\pi\sin (2\theta)}
-\frac{2\cos (2\theta)}{\pi}+\frac{\pi}{2}\bigg]^{-1},
\label{EC47}
\end{aligned}
\end{equation}

\noindent
where the limits $\theta_0$ and $\theta_c$ correspond to the critical
angles for strain $\varepsilon =0$ and the final value $\varepsilon$
of the strain, respectively.

\begin{figure}[h!]
\begin{center}
\includegraphics[width=8cm]{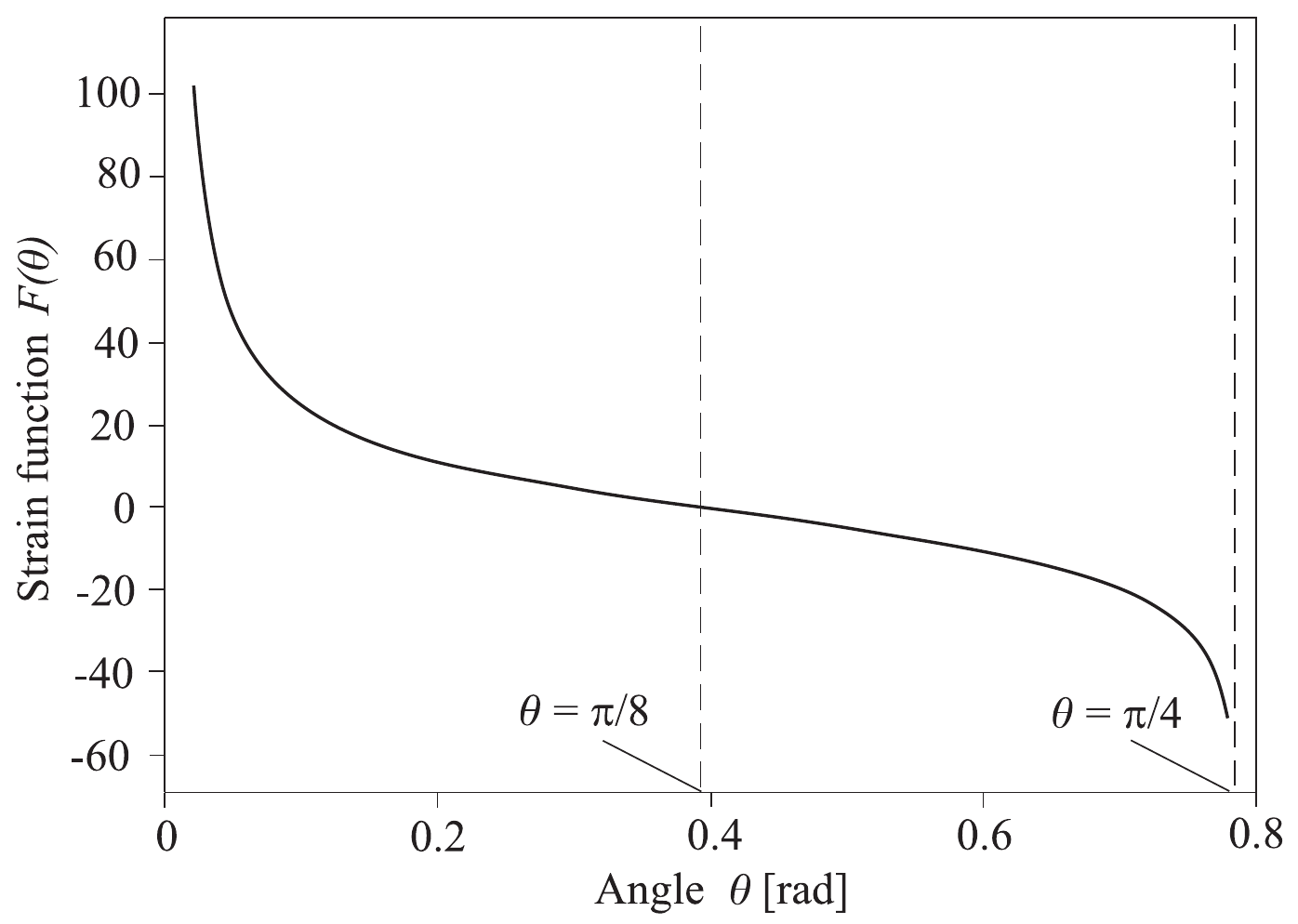}
\caption{\label{Fig10} The strain universal function $F(\theta)$ and
its asymptotas at $\theta =0$ and $\theta =\pi /4$.}
\end{center}
\end{figure}

Hence the strain $\varepsilon$ is related with the auxiliary variable
$\theta_c$ by an expression of the form

\begin{equation}
\varepsilon =\frac{k_BT}{B\Omega^*}[F(\theta_c)-F(\theta_0)],
\label{EC48}
\end{equation}

\noindent
where $F(\theta)$ is the universal function

\begin{equation}
\begin{aligned}
F(\theta)&=-2\int d\theta\, \cot (2\theta)\sin (2\theta)
\bigg[( 1-\cos (2\theta)\\
&-2\theta\sin (2\theta)
-\frac{4\theta}{\pi}-\frac{\sin (4\theta)}{\pi}
+\frac{\pi\sin (2\theta)}{2}\bigg]^{-1},
\label{EC49}
\end{aligned}
\end{equation}

\noindent
whose graph is depicted in Fig.~\ref{Fig10}. $F(\theta)$ is
monotonically decreasing in its whole range $(0,\pi/4)$, and diverges
at $\theta =0$ and $\theta =\pi/4$.

The equation for the critical angle $\theta_0$ at $\varepsilon =0$
follows from combining Eqs.~(\ref{EC38}) and (\ref{EC39}) in order to
eliminate $\sigma$. It is obtained

\begin{equation}
\begin{aligned}
\dot\varepsilon =2C_0\frac{\Omega^*\tau_c}{k_BT}
&\left[\cot (2\theta_0)+2\theta_0-\frac{\pi}{2}\right]\\
&\times\exp\left[-\frac{1}{k_BT}\left(\epsilon_0-
\frac{2\Omega^*\tau_c}{3\sin (2\theta_0)}\right)\right].
\label{EC50}
\end{aligned}
\end{equation} 

\noindent
This equation gives $\theta_0$ as a function of $\dot\varepsilon$ and
$T$. Therefore, the set of Eqs.~(\ref{EC48}), (\ref{EC49}) and
(\ref{EC50}) allows to determine $\theta_c=\theta_c(\varepsilon,\,
\dot\varepsilon ,\, T)$.    

The adimensional coefficient appearing in Eq.~(\ref{EC48}) establishes
the scale for $\varepsilon$ and is a very small number. The bulk
elastic modulus $B$ is close to $0.7\times 10^{11}\,\text{Pa}$ for
aluminium and $1.0\times 10^{11}\,\text{Pa}$ for titanium at normal
room temperature. From the comparison of our curves with experiments
for $\varepsilon\approx 0$, we know that $\Omega^*$ is $2.6\times
10^{-21}\,\text{cm}^3$ for Al--8090 and $5.9\times
10^{-22}\,\text{cm}^3$ for Ti--6Al--4V at rather high temperatures,
however, $\Omega^*$ is not expected to vary with $T$.  Hence, at
$T=300\,\text{K}$,

\begin{equation}
\frac{k_BT}{B\Omega^*}\thicksim
2.3\times 10^{-5}\, -\, 7.0\times 10^{-5}.
\label{EC51}
\end{equation}

\noindent   
At elevated temperatures, $B$ may decrease substantially, but anyway
the coefficient (\ref{EC51}) is expected to remain small. 

Because of the small value of the coefficient (\ref{EC51}), any
significative strain $\varepsilon$ demands that either $\theta_0$ or
$\theta_c$, or both, be in one of the two asymptotic regions
$\theta\gtrsim 0$ or $\theta\lesssim\pi/4$. The asymptotic region
$\theta\lesssim\pi/4$ can be associated to creep, while $\theta\gtrsim
0$ is related to ductile and superplastic deformation. For example, to
attain a strain $\varepsilon\thicksim 1-10\,\%$ it must be
$F(\theta_c)-F(\theta_0)\thicksim 1000$. Observing Fig.~\ref{Fig10}
one readily realizes that such big differences can only exist in the
asymptotic regions of $F(\theta)$.

\subsection{The strain and strain rate dependent stresses}
\label{straindependentstresses}

Combining Eqs.~(\ref{EC33}) and (\ref{EC35}) to eliminate
$\mathcal{Q}$, and then isolating $p$, one obtains the general
equation

\begin{equation}
\begin{aligned}
p=\frac{k_BT}{\Omega^*}&\bigg[ \ln\left(
\frac{2C_0\Omega^*\tau_c}{k_BT\dot\varepsilon}\right)\\
&+\ln\left( \cot (2\theta_c)+2\theta_c-\frac{\pi}{2}\right)\bigg]
-\frac{\epsilon_0}{\Omega^*},
\label{EC52}
\end{aligned}
\end{equation}

\noindent
which gives the pressure once the relation $\theta_c=
\theta_c(\varepsilon,\,\dot\varepsilon ,\, T)$ is known from the
procedure explained in the previous subsection.

Knowing $\theta_c=\theta_c(\varepsilon,\,\dot\varepsilon ,\, T)$ and
$p=p(\varepsilon,\,\dot\varepsilon ,\, T)$, the tensile stress
$\sigma$ can be obtained from Eq.~(\ref{EC33a}), which can be written
as

\begin{equation}
\sigma =\frac{4\tau_c}{3\sin (2\theta_c)}-p
\label{EC53}
\end{equation}

\noindent
or, more explicitly,

\begin{equation}
\begin{aligned}
\sigma =\frac{4\tau_c}{3\sin (2\theta_c)}
&-\frac{k_BT}{\Omega^*}\bigg[ \ln\left(
\frac{2C_0\Omega^*\tau_c}{k_BT\dot\varepsilon}\right)\\
&+\ln\left( \cot (2\theta_c)+2\theta_c-\frac{\pi}{2}\right)\bigg]
+\frac{\epsilon_0}{\Omega^*}.
\label{EC54}
\end{aligned}
\end{equation}

\noindent
The transversal compression $\sigma_x=\sigma_y$ follows from
Eqs.~(\ref{EC28a}).

\subsection{The constitutive equation for $\sigma\gg\tau_c$}
\label{largesigma}

The procedure outlined in the preceding subsection for determining the
exact dependence of the flow stress on strain, strain rate and
temperature relies on solving a set of trascendental
equations. However, a rather simple closed--form constitutive equation
$\sigma =\sigma(\varepsilon ,\dot\varepsilon,T)$ can be written if one
finds acceptable to restrict the scope to the strain rates for which
Eq.~(\ref{EC40}) is valid. This means $\dot\varepsilon >2\times
10^{-4}\,\text{s}^{-1}$ for Ti--6Al--4V, which is a typical figure of
more or less general validity.

The coefficient $k_BT/(B\Omega^*)$ appearing in Eq.~(\ref{EC48}) is in
general a very small adimensional quantity. As $F(\theta)$ is an
universal function exhibiting large variations only close to its
sigularities at $\theta =0$ and $\theta =\pi/4$, significant strains
$\varepsilon$ demand $\theta_c\gtrsim 0$ or $\theta_c\lesssim\pi/4$.
In the latter eventuality the stress is near the minimum necessary to
produce plastic flow, which progresses at a very low rate and
corresponds to {\it creep}. The former case is the ductile deformation
we are interested in.

When $\theta_c\gtrsim 0$ the universal function $F(\theta)$, given by
Eq.~(\ref{EC49}), can be substituted by the asymptotic equation

\begin{equation}
F(\theta)=\frac{1}{\pi -8/\pi}\frac{1}{\theta}
\label{ED1}
\end{equation}

\noindent
and then the expression (\ref{EC48}) for the strain becomes

\begin{equation}
\varepsilon =
\frac{1}{\pi -8/\pi}\frac{k_BT}{B\Omega^*}\frac{1}{\theta_c}
+\varepsilon_0 ,
\label{ED2}
\end{equation}

\noindent
where $\varepsilon_0$ is an integration constant determined by the
condition (\ref{EC34}). Isolating $\theta_c$ from Eq.~(\ref{ED2}) and
inserting in Eqs.~(\ref{EC52}) and (\ref{EC53}), both rewritten for
$\theta_c\ll 1$, it gives

\begin{equation}
p=\frac{k_BT}{\Omega^*}
\ln\left[\frac{(\pi -8/\pi)C_0(\Omega^*)^2B\tau_c}
{(k_BT)^2\dot\varepsilon}
(\varepsilon +\varepsilon_0)\right] -\frac{\epsilon_0}{\Omega^*}
\label{ED3}
\end{equation}

\noindent
and

\begin{equation}
\sigma=\frac{2(\pi -8/\pi)}{3}\frac{B\Omega^*\tau_c}{k_BT}
(\varepsilon +\varepsilon_0)-p.
\label{ED4}
\end{equation}

\noindent
The initial condition (\ref{EC34}), that is, $p=-\sigma_0/3$ when
$\varepsilon =0$, with $\sigma_0$ being the flow stress for
$\varepsilon =0$, allows us to put Eqs.~(\ref{ED3}) and (\ref{ED4}) as

\begin{equation}
p=\frac{k_BT}{\Omega^*}
\ln\left[\frac{C_0\Omega^*}{k_BT\dot\varepsilon}
\left( \frac{(\pi-8/\pi)B\Omega^*\tau_c}{k_BT}\,\varepsilon
+\sigma_0\right)\right] -\frac{\epsilon_0}{\Omega^*}
\label{ED5}
\end{equation}

\noindent
and

\begin{equation}
\begin{aligned}
\sigma&=\frac{2(\pi -8/\pi)}{3}\frac{B\Omega^*\tau_c}{k_BT}
\,\varepsilon +\frac{2}{3}\sigma_0 \\
&-\frac{k_BT}{\Omega^*}
\ln\left[\frac{C_0\Omega^*}{k_BT\dot\varepsilon}
\left( \frac{(\pi-8/\pi)B\Omega^*\tau_c}{k_BT}\,\varepsilon
+\sigma_0\right)\right] +\frac{\epsilon_0}{\Omega^*},
\label{ED6}
\end{aligned}
\end{equation}

\noindent
which is the general constitutive equation $\sigma =\sigma(\varepsilon
,\dot\varepsilon,T)$ we were searching for.  If one replaces in
Eq.~(\ref{ED6}) $\varepsilon=0$ and $\sigma =\sigma_0$, the
approximate constitutive equation (\ref{EC40}) for $\varepsilon=0$ is
recovered. The domain in which the approximation holds is apparent in
Fig.~\ref{Fig7}, where the curves given by Eq.~(\ref{EC40}) and the
exact Eq.~(\ref{EC39}) are compared.

\subsection{Exact expression for the slope $\partial\sigma
/\partial\varepsilon$ at $\varepsilon =0$}
\label{slope}

Derivating and combining properly the equations written in subsections
\ref{strainfunction} and \ref{straindependentstresses} one can deduce
a closed--form equation for the slope
$\partial\sigma /\partial\varepsilon$ at $\varepsilon =0$. In general,
for any $\varepsilon$,

\begin{equation}
\begin{aligned}
\frac{\partial\sigma}{\partial\varepsilon} = 
&B\,\bigg( \frac{4\Omega^*\tau_c}{3k_BT\cos(2\theta_c)}
-\frac{1}{\cot(2\theta_c)+2\theta_c-\pi/2}\bigg) \\
&\times\bigg[\frac{1-\cos (2\theta_c)}{\sin (2\theta_c)}
-2\theta_c\bigg( 1+\frac{2}{\pi\sin (2\theta_c)}\bigg) \\
&-\frac{2\cos (2\theta_c)}{\pi}+\frac{\pi}{2}\bigg].
\end{aligned}
\label{ER23}
\end{equation}

\noindent
For $\varepsilon =0$ one can replace $\sin
(2\theta_c)=2\tau_c/\sigma$, and Eq.~(\ref{ER23}) becomes an
expression of only the initial value of $\sigma$.

The slope $\partial\sigma /\partial\varepsilon$ at $\varepsilon =0$ is
a very important quantity because it is observed that always the
maximal superplastic strain to failure is attained close to a
temperature and strain rate such that

\begin{equation}
\frac{\partial\sigma}{\partial\varepsilon}\bigg|_{\varepsilon =0}=0,
\label{ER24}
\end{equation}

\noindent
which may be considered as an analytical condition for
superplasticity.  Although this conclusion is mostly empirical, it is
quite reasonable because, as a polycrystal cannot flow steadily by the
monotonic increase of the internal pressure $p$, Eq.~(\ref{ER24})
expresses the situation which best emulate a steady regime. The next
example illustrates the point.

\begin{figure}[h!]
\begin{center}
\includegraphics[width=7cm]{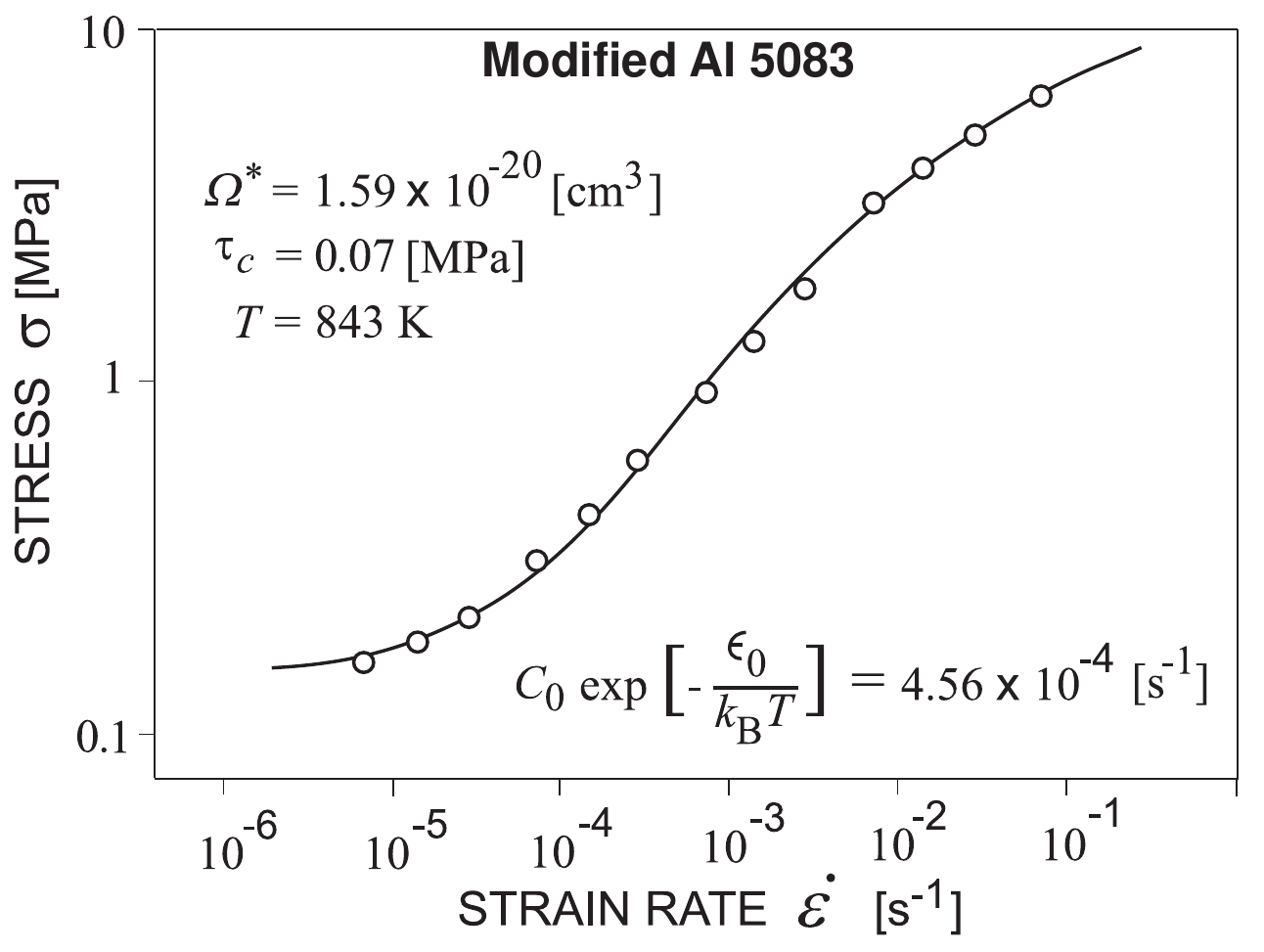}
\caption{\label{Fig11} The uniaxial flow stress $\sigma$ for a
modified Al 5083 alloy deformed at $570\,^\circ\text{C}$ and constant
strain rate $\dot\varepsilon$. Circles represent the data of Kaibyshev
{\it et.~al.} (Ref.~\onlinecite{Kaibyshev}) and solid line depicts
Eq.~(\ref{EC39}) with the material constants shown in the insets.}
\end{center}
\end{figure}

\begin{figure}[h!]
\begin{center}
\includegraphics[width=7cm]{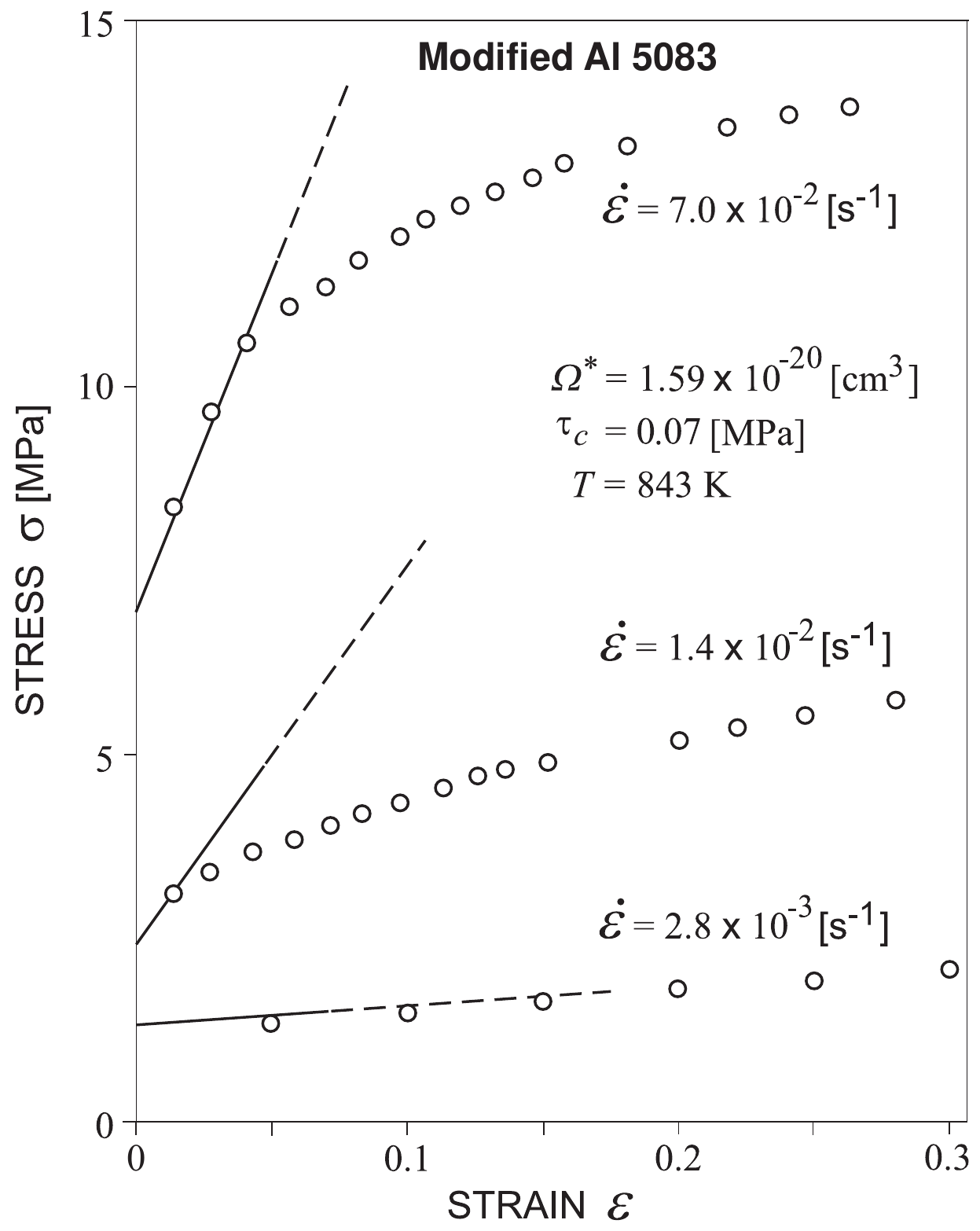}
\caption{\label{Fig12} Circles represent stress--strain data at
constant strain rates $\dot\varepsilon =2.8\times 10^{-3},\,1.4\times
10^{-2}$ and $7.0\times 10^{-2}\,\text{s}^{-1}$ for the same system of
Fig.~\ref{Fig11}. The straight lines are the tangents to the
stress--strain curves at $\varepsilon =0$ and the same strain rates,
calculated with Eq.~(\ref{ER23}). The effective bulk modulus is
$B=3000\,\text{MPa}$.}
\end{center}
\end{figure}

Fig.~\ref{Fig11} shows the variation of flow stress with strain rate
for a modified 5083 aluminium alloy, deformed at $T=843\,\text{K}$ and
constant true strain rate. The experimental points are due to
Kaibyshev {\it et.~al.}~\cite{Kaibyshev} and the solid line depicts
Eq.~(\ref{EC39}) with the constants shown in the insets. These authors
observed a sharp maximum in elongation to failure of 1150\% for
$T=843\,\text{K}$ and a strain rate $\dot\varepsilon
=2.8\times10^{-3}\,\text{s}^{-1}$.

Inserting in Eq.~(\ref{ER23}) the constants obtained from the fit of
the stress--strain rate data, shown in Fig.~\ref{Fig11}, the slope
$\partial\sigma /\partial\varepsilon$ of the stress--strain curves at
$\varepsilon =0$ can be calculated for any initial stress
$\sigma$. The solid--dashed straight lines drawn in Fig.~\ref{Fig12}
represent the calculated tangents to the stress--strain curves at
$\varepsilon =0$ starting from $\sigma =1.33,\,2.4$, and
$7.0\,\text{MPa}$. According to Eq.~(\ref{EC39}) these three stresses
correspond to the strain rates $\dot\varepsilon =2.8\times
10^{-3},\,1.4\times 10^{-2}$ and $7.0\times 10^{-2}\,\text{s}^{-1}$.
The circles in Fig.~\ref{Fig12} represent the experimental
stress--strain data taken by Kaibyshev {\it et.~al.} at constant
strain rates $\dot\varepsilon =2.8\times 10^{-3},\,1.4\times 10^{-2}$
and $7.0\times 10^{-2}\,\text{s}^{-1}$. The consistency between theory
and experiment is apparent in Fig.~\ref{Fig12}. The effective bulk
modulus, which fixes the scale for $\varepsilon$, was taken as
$B=3000\,\text{MPa}$. Best superplastic response of the material
is attained at a strain rate $\dot\varepsilon =2.8\times 10^{-3}$,
for which $\partial\sigma /\partial\varepsilon\approx 0$ at
$\varepsilon =0$. 

\subsection{Grain sliding and the sign of the concavity of the
stress--strain curve} \label{concavity}

In a recently published experimental study of the plastic tensile
deformation of Ti--6Al--4V, Vanderhasten {\it et.~al.}~find out that
positive concavity seems to be the fingerprint of grain boundary
sliding \cite{Vanderhasten1, Vanderhasten2}. These authors conducted
tests starting from room temperature to 1323 K, recording tensile
properties, microstructural evolution and crystallographic texture
changes at a constant true strain rate $\dot\varepsilon =5\times
10^{-4}\,\text{s}^{-1}$. The tests seem not intended to attain optimal
superplastic performance, but clearly show a dramatic increase of the
strain to fracture at $T=873\,\text{K}$ ($600^\circ\,\text{C})$, which
reaches the optimum ($\varepsilon =1.6$) at $T=1123\,\text{K}$
($850^\circ\,\text{C}$). Starting from $T=998\,\text{K}$, grain
boundary sliding begins to be dominant and, coincidently, the
corresponding stress--strain curve exhibits positive concavity over
the most of the strain domain. The lower temperature range is
characterized by negative concavity and grain stretching. At
temperatures between $T=1123\,\text{K}$ and $T=1173\,\text{K}$ the
stress--strain curves have positive concavity over the whole range,
from $\varepsilon =0$ to fracture at $\varepsilon\approx 1.6$. For
higher temperatures the strain to fracture decreases and the
stress--strain curves develop a final sector with small negative
curvature before failure.

\begin{figure}[h!]
\begin{center}
\includegraphics[width=8cm]{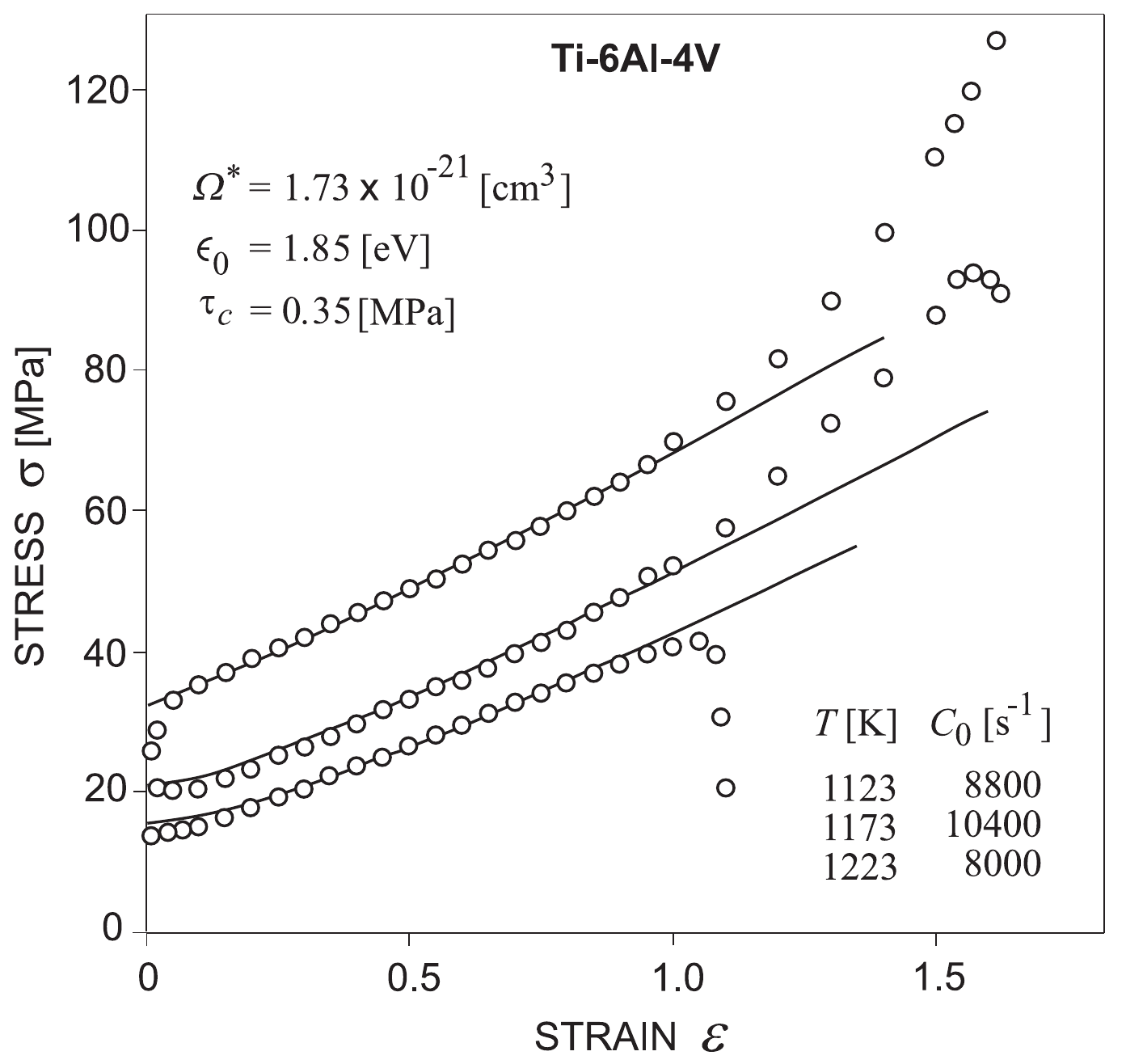}
\caption{\label{Fig13} Circles represent data for Ti--6Al--4V of
Vanderhasten {\it et.~al.} at the the temperatures for which grain
sliding dominates over grain stretching. Solid lines were obtained
from Eq.~(\ref{ED6}) and the same set of parameters, which are shown
in the insets. Just $C_0$ shows some variation with $T$, probably due
to its grain size dependence. The effective bulk modulus is
$B=3000\,\text{MPa}$.}
\end{center}
\end{figure}

\begin{figure}[h!]
\begin{center}
\includegraphics[width=8cm]{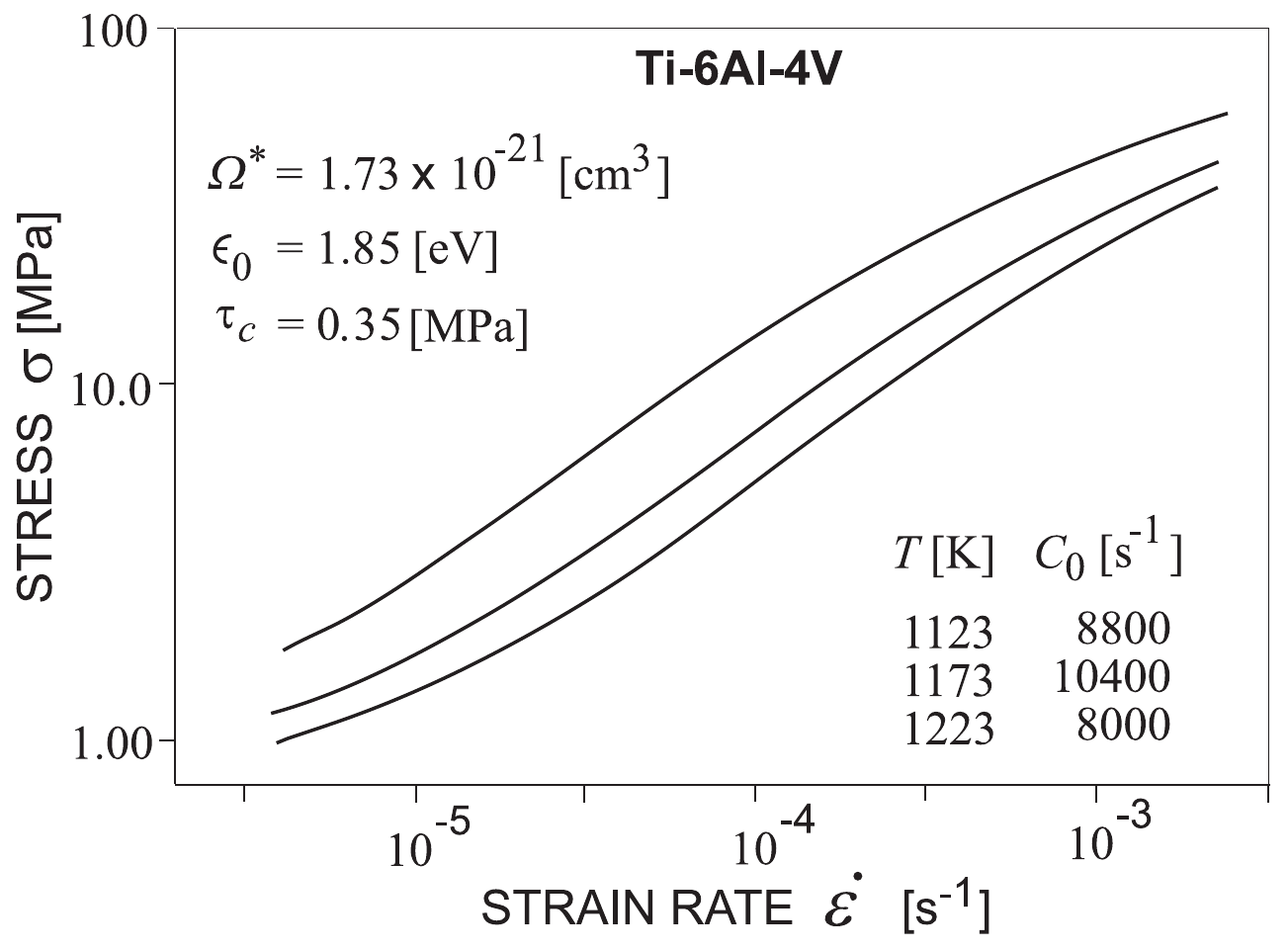}
\caption{\label{Fig14} Stress--strain rate curves calculated from
Eq.~(\ref{EC39}) and the same material constants of Fig.~\ref{Fig13}.
No experimental data are available to compare with.}
\end{center}
\end{figure}

The open circles in Fig.~\ref{Fig13} represent experimental data taken
by Vanderhasten {\it et.~al.}~on the stress--strain response of the
titanium alloy Ti--6Al--4V at the three temperatures for which the
deformation is observed to take place by grain sliding
\cite{Vanderhasten1,Vanderhasten2}.  The strain rate was kept at
$\dot\varepsilon =5.0\times 10^{-4}\,\text{s}^{-1}$ in all runs. The
samples exhibit high ductility but not properly superplasticity. The
solid lines represent the asymptotic expression (\ref{ED6}) for the
three temperatures, with the parameters shown in the insets and
$B=3000\,\text{MPa}$. Eq.~(\ref{ED6}) can describe with high precision
the experimental situation, which can be shown by comparing the values
for $\sigma$ at $\varepsilon =0$ given by Eq.~(\ref{ED6}) with the
exact value $\sigma_0$, obtained from Eq.~(\ref{EC39}) to introduce it
into the same Eq.~(\ref{ED6}) as a one of its constants. The agreement
between theory and experiment is remarkable up to $\varepsilon =1$.

Fig.~\ref{Fig14} shows the stress--strain rate curves given by
Eq.~(\ref{EC39}) and the same material constants used to fit the data
of Fig.~\ref{Fig13}. There is no experimental points to compare with
since Vanderhasten {\it et.~al.}~employed a unique strain rate. The
curves exhibit the typical sigmoidal shape.

\begin{figure}[h!]
\begin{center}
\includegraphics[width=8cm]{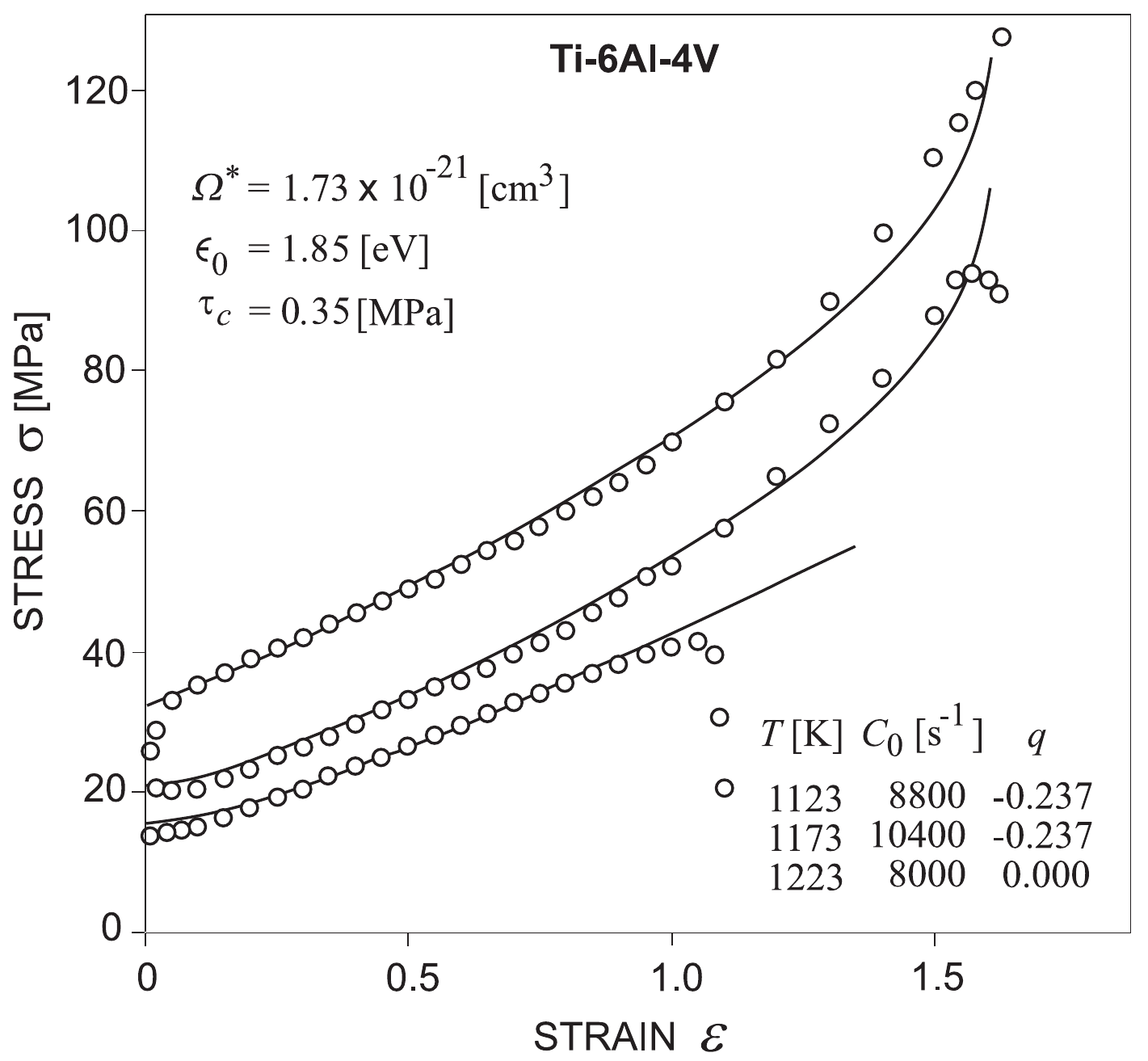}
\caption{\label{Fig15} Same as Fig.~\ref{Fig13}, but with $C_0$
replaced by $C_0^\prime$, as given by Eq.~(\ref{ER25}) with the values
of $q$ shown in the inset. $C_0^\prime$ models the effect of grain
size variations.}
\end{center}
\end{figure}

In general, $C_0$ depends on the grain size $d$ as $d^{-3}$ and the
rest of the parameters do not vary with $d$
\cite{Lagos2}. Fig.~\ref{Fig15} shows an attempt to explain the
departure between theory and experiment, starting at $\varepsilon =1$,
by the dynamic modification of the grain size. The curves represent
Eq.~(\ref{ED6}) with $C_0$ substituted by

\begin{equation}
C_0^\prime =(1+q\varepsilon^3)C_0, 
\label{ER25}
\end{equation}

\noindent
where $q$ is a constant. The values of $q$ giving the best fit to
experiment are shown in one of the insets of Fig.~\ref{Fig15}.
Amazingly, $q$ turns out to be such that $C_0^\prime =0$ at the
fracture strain. This means that failure occurs because the grains
become unable to slide, which can hardly be explained by just their
size variations. The explanation may be as follows: The mean grain
sizes measured by Vanderhasten {\it et.~al.}~at 1123 and 1173 K show a
rather irregular reduction on strain. The consequent division of old
grains involves the creation of new grain surfaces which do not have
necessarily the composition and structure of the old ones. Grain
boundaries are efficient traps for impurities, and can condense atomic
species having a low bulk concentration. However, the process is
governed by bulk diffusion and is not rapid. Therefore, concerning
their composition and consequent sliding ability, fresh grains may be
very different than well--aged ones. From this viewpoint, grain size
variations may simply reduce the number of grains able to slide. This
explains also why the smooth stress--strain curves do not reflect the
irregularities observed in the grain size--strain curves.

The theoretical approach to finite strain superplasticity put forward
here permits to explain quantitatively the main features observed in
the superplastic deformation. However, the theory lacks an important
element: the effect of heat generation at the sliding grain
boundaries. In our scheme, the whole work made by the external forces
is dissipated into heat at the sliding grain boundaries, which occupy
a very small volume when compared with the grain bulk. Hence one may
expect that grain boundaries will be at a higher temperature than bulk
matter, being the nominal temperature the bulk one. Grain boundaries
whose normals are within solid angles, determined by $\theta_c$,
around the principal stress directions do not slide because the
in--plane shear stress is smaller than $\tau_c$, and then remain at
the bulk temperature. Bending downward of the stress--strain curves at
high strains and strain rates seems to be a grain boundary temperature
effect.

\subsection{Two phase alloys and the nature of the superplastic
deformation} \label{twophase}

The scheme described thus far should hold for single phase
alloys. Materials combining two or more allotropic phases in
thermodynamic equilibrium may display a more complex phenomenology.
In particular, superplastic materials are generally two--phase or
pseudo single phase alloys. Probably the most characteristic example
of the former is Ti--6Al--4V, which combines an $\alpha$ hcp solid
phase with a $\beta$ phase with bcc structure, in comparable volume
fractions up to the $\beta$ transus temperature. Cope el
al. \cite{Cope2, Cope3} observed best superplastic response of
Ti--6Al--4V for a temperature $T=1153\,\text{K}$, for which the
$\beta$ phase proportion was 0.42\%. In pseudo single phase materials
the second phase occurs as small inclusions between the grains
constituting the main one.

The new feature introduced by the coexistence of two allotropic phases
in thermodynamical equilibrium is that a volume reduction is not
necessarily related with just the elastic properties of the
material. If the two phases (which we will call generically as
$\alpha$ and $\beta$) have different specific volumes, the total
volume may change also by the transformation of part of one of the
phases into the other one. In general, the atomic fraction of, for
instance, the $\beta$ phase

\begin{equation}
\frac{N_\beta}{N}=\eta_\beta (T,p),
\label{EC55}
\end{equation}

\noindent
where $N_\beta$ is the number of atoms of the main constituent of the
alloy in the $\beta$ phase and $N=N_{\alpha}+N_{\beta}$ is the total,
is a funtion of the temperature $T$ and pressure $p$. If $p$ changes
with time, the atomic fraction of the phases changes, as well as the
total volume $V$. Hence, the dilation rate is no more given by
Eq.~(\ref{EC30}) but by

\begin{equation}
\frac{\dot V}{V}=
\left[ -\frac{1}{B}+\frac{v_\beta -v_\alpha}{N_\alpha v_\alpha
+N_\beta v_\beta}
\left( \frac{\partial N_\beta}{\partial p} \right)_T \right]\,\dot p,
\label{EC56}
\end{equation}

\noindent
where $v_\alpha$ and $v_\beta$ are the volume per atom of the two
phases.

Eq.~(\ref{EC56}) assumes that the atomic fractions have their
equilibrium values in any instant, and hence a quasi--static process,
otherwise the transformation kinetics would come into the fore. The
total area of the interfaces between the phases is maximal when their
volumes have the same value, and then comparable volume fractions
contribute to increase the phase transformation rates, and the
accuracy of Eq.~(\ref{EC56}). Also, the new isothermal coefficient
appearing in Eq.~(\ref{EC56}) is in general not a constant, and it is
expected to be much more sensitive to temperature than $1/B$. With
this in mind we define

\begin{equation}
\frac{1}{B_{\alpha\beta}}=-\frac{v_\beta -v_\alpha}{N_\alpha v_\alpha
+N_\beta v_\beta}
\left(\frac{\partial N_\beta}{\partial p}\right)_T 
\label{EC57}
\end{equation}

\noindent
and call

\begin{equation}
\frac{1}{B^*}=\frac{1}{B}+\frac{1}{B_{\alpha\beta}}.
\label{EC58}
\end{equation}

\noindent
$B_{\alpha\beta}$ is a positive quantity since the two factors in the
right hand side of Eq.~(\ref{EC57}) have opposite signs, as demanded
by mechanical stability.

The equations of the previous sections are formally recovered this
way, with an effective bulk modulus $B^*$ instead of the purely
elastic one $B$.

The elastic modulus $B$ takes in general large values (typically
$B\approx 1\times 10^5\,\text{MPa}$ for metals), which is not
necessarily the case of $B_{\alpha\beta}$. For two phases coexisting
in thermodynamical equilibrium one may expect that the volume fraction
of each one would vary significantly for pressure variations much
smaller than the yield stress. Then, values of less than 1 MPa for
$B_{\alpha\beta}$ are quite reasonable. In such a situation $1/B\ll
1/B_{\alpha\beta}$ and $1/B^*$ is several orders of magnitude larger
than $1/B$. By this reason solids exhibiting two allotropic phases in
equilibrium may display a special plastic behaviour, of which
superplasticity seems to be an example.

The abnormally large strains to failure displayed by superplastic
materials have been attributed to the better stability they have
against necking in tensile tests. The plastic flow can be represented
by the relation

\begin{equation}
\sigma =K\dot\varepsilon^m ,
\label{EC59}
\end{equation}

\noindent 
where $m$ is a function of $\dot\varepsilon$ and $\varepsilon$,
generally known as the strain rate sensitivity, which is always
smaller than unity. When $m=1$ any irregularity in the cross section
of the sample is not accentuated by the deformation in a tensile test,
necking has no effect and the material behaves as a Newtonian viscous
fluid.  Necking instability is high for $m$ below 0.5 and decreases
markedly when $m\ge 0.5$. Indeed, the condition of high values of $m$
is a necessary requirement, but not enough to reach elongations as
large as those termed superplastic. Our previous discussions show that
plastic elongation has a limit because demands a volume reduction, and
failure is not a matter of just a mechanical instability of geometric
origin.

Therefore, in our view, superplasticity has its real origin in three
concurrent features of the material and physical conditions of the
deformation process: (a) A high strain sensitivity $m$. (b) The
ability of the material to reduce the internal pressure $p$ by
undergoing a phase transformation to a phase of higher density. (c)
The material must be in the verge of a ductile to brittle
transition. This latter requirement is explained below, taking the
superplastic behaviour of titanium Ti--6Al--4V to illustrate the
argument.

\subsection{The high temperature second yield point}
\label{instability}

The general properties of the stress {\it vs.}~strain curves can be
visualized from just examining Eq.~(\ref{ED6}) for the strain
dependent flow stress $\sigma$. It has a first term linear in
$\varepsilon$, with slope proportional to $1/T$, and a negative
logarithmic term whose coefficient increses with $T$. Hence, at low
temperatures $\sigma$ is linear in $\varepsilon$ and, as $T$
increases, the slope decreases and the function develops positive
concavity. Beyond a critical temperature for which the slope at the
origin vanishes the curve starts at $\varepsilon =0$ whith negative
slope and, after surpassing a minimum, or second yield point, begins
to increase monotonically.

Hence, Eq.~(\ref{ED6}) predicts strain hardening at any strain for
temperatures below a critical temperature, which roughly coincides
with the temperature which optimizes superplastic properties. At
higher temperatures the material should weaken for strains between
$\varepsilon =0$ and the second yield point. This configures a
mechanically unstable situation which cannot occur in practice in a
smooth and homogeneous way. The equations written so far describe the
mechanical response at a point, which does not necessarily can be
straightforwardly extended to the whole sample. Homogeneity is a quite
reasonable assumption to study averaged properties in the macroscopic
scale, but may fail in critical situations. We can safely assume that
$\Omega^*$ and $\epsilon_0$ are uniform and constant throughout the
material, but $C_0$ depends on the grain size as $d^{-3}$
\cite{Lagos2}, and hence may fluctuate or have a multimodal
distribution, following grain size.

There is a class of mild steels \cite{Song} and aluminium alloys
\cite{Hayes, Wang} exhibiting stress--strain curves with two yield
points.  The plastic deformation of these materials is inhomogeneous
and takes place by a series of succesive sudden band formations
(L\"uders bands
\cite{Hutanu}), transversal to the tensile axis and making a rather
precise angle with it (of about $56^\circ$), which concentrate the
strain. The flow stress keeps constant (though displaying a serrated
aspect) until the bands, increasing in number and extension, fill the
whole sample and the complete material becomes strained beyond the
second yield point strain. Just then the stress starts to increase. In
the intermediate steps, samples have sections with no strain and
sections strained beyond the second yield strain. The relevant
question is then how a system with the stress--strain curves given by
Eq.~(\ref{ED6}) will evolve with strain, and what kind of effective
stress--strain curves could be expected from such evolution.

If our material is not exactly homogeneous the critical temperature
for the onset of the second yield point is not the same in its whole
extension. Hence, at temperatures close to these critical temperatures
and small strain one may expect the solid will be partitioned in a
network of sectors hardening on strain, and another one which
weakens. As the stress is uniform, the weakening sectors will deform
much faster, and will be responsible for the main part of the
effective strain before reaching their second yield point. Meanwhile
the hardening sectors will furnish mechanical stability against sudden
yield steps, like L\"uders band generation.

Anyway, the condition for superplastic behaviour seems related with
the onset of the second yield point, producing a transition to a
potentially unstable system. Such a transition begins at the
temperature for which $\partial\sigma /\partial\varepsilon$ at
$\varepsilon =0$ vanishes and changes from positive to negative
sign. On the other hand, generally speaking, the most basic condition
for a steady flow, as superplastic deformation tends to be, is that
the dynamical variables remain constant and independent of
$\varepsilon$, which here represents time. As $\sigma$ varies with
$\varepsilon$, all what can be done is to emulate a steady flow by
providing the conditions that keep $\partial\sigma
/\partial\varepsilon$ small and steady in a range as large as
possible, which incorporates $\varepsilon =0$.

\begin{acknowledgments}
The authors thanks the support by the Programa Bicentenario de Ciencia
y Tecnolog\'ia (Chile) ACT--26.
\end{acknowledgments}


\begin{thebibliography}{10}

\bibitem{Pilling&Ridley} J. Pilling and N. Ridley, {\it
Superplasticity in crystalline solids}, (The Institute of Metals,
Camelot, Southampton, UK, 1989).

\bibitem{Nieh} T. G. Nieh, J. Wadsworth and O. D. Sherby, {\it
Superplasticity in metals and ceramics} (Cambrige, UK 1997).

\bibitem{Cope2} M. T. Cope and N. Ridley, Mat. Sci. Technol. {\bf 2},
140 (1986).

\bibitem{Backofen} T. H. Thomsen, D. L. Holt and W. A. Backofen,
Met. Eng. Quart. {\bf 2}, 1 (1970).

\bibitem{AerospaceMat} {\it Aerospace Materials}, edited by B. Cantor,
H. Assender and P. Grant. (Institute of Physics Publishing, Bristol,
UK, 2001).

\bibitem{Khaleel} M. A. Khaleel, H. M. Zbib and E. A. Nyberg,
Int. J. Plast. {\bf 17}, 277 (2001).

\bibitem{Taylor}M. B. Taylor, H. M. Zbib and M. A. Khaleel,
Int. J. Plast. {\bf 18}, 415 (2002).

\bibitem{Livesey1} D. W. Livesey and N. Ridley, Metall. Trans.
{\bf 9A}, 519 (1978).

\bibitem{Livesey2} D. W. Livesey and N. Ridley, Metall. Trans.
{\bf 13A}, 1619 (1982).

\bibitem{Cope3} M. T. Cope, D. R. Evetts and N. Ridley,
J. Mat. Sci. {\bf 21}, 4003 (1986).

\bibitem{Ball} A. Ball and M. M. Hutchinson, Met. Sci. J. {\bf 3},
1,(1969).

\bibitem{Mukherjee1} A. K. Mukherjee, Mater. Sci. Eng. {\bf
8}, 83 (1971).  

\bibitem{Langdon} T. G. Langdon, Phil. Mag., {\bf
22A}, 689 (1970).

\bibitem{Hayden} H. W. Hayden, S. Floreen and P. D. Goodall,
Metall. Trans. {\bf 3A}, 833 (1972).

\bibitem{AshbyVerrall} M. F. Ashby and R. A. Verrall, Acta Metall.
{\bf 21}, 149 (1973).

\bibitem{Gifkins} R. C. Gifkins, Metall. Trans. {\bf 7A}, 1225 (1976).

\bibitem{Mukherjee2} A. Arieli and A. K. Mukherjee, Mater. Sci. Eng.
 {\bf 45}, 61 (1980).

\bibitem{Fukuyo} H. Fukuyo, H. C. Tsai, T. Oyama and O. D. Sherby,
ISIJ International {\bf 31}, 76 (1991).

\bibitem{WangFu} C. G. Wang, M. W. Fu, C. X. Cao and H. B. Dong,
Mat. Sci. Eng. A, in press. 

\bibitem{Lagos0}  M. Lagos and H. Duque, Int. J. Plast. {\bf 17}, 369
(2001).

\bibitem{Lagos1} M. Lagos, Phys. Rev. Lett. {\bf 85}, 2332 (2000).

\bibitem{Lagos2} M. Lagos, Phys. Rev. B {\bf 71}, 224117 (2005).

\bibitem{Lagos3} M. Lagos, Phys. Rev. B {\bf 73}, 224107 (2006).

\bibitem{Bellon} P. Bellon and R. S. Averback, Phys. Rev. Lett. {\bf
74}, 1819 (1995).

\bibitem{Srolovitz} D. J. Srolovitz, Acta Metall. {\bf 37}, 621 (1989).

\bibitem{Muller} J. M\"uller and M. Grant, Phys. Rev. Lett. {\bf 82},
1736 (1999).

\bibitem{Grinfeld} M. A. Grinfeld, Dokl. Acad. Nauk. SSSR {\bf 265},
836 (1982); {\sl Sov. Phys. Dokl. \bf 31}, 831 (1986); Europhys.
Lett. {\bf 22}, 723 (1993).

\bibitem{Williams} D. R. M. Williams, Phys. Rev. Lett. {\bf 75}, 453
(1995).

\bibitem{Genin} F. Y. G\'enin, J. Appl. Phys. {\bf 77}, 5130 (1995).

\bibitem{Jones} D. E. Jones, J. P. Pelz, Y. Hong, E. Bauer and I. S.
T. Tsong, Phys. Rev. Lett. {\bf 77}, 330 (1996).

\bibitem{Jesson} D. E. Jesson, K. M. Chen, S. J. Pennycook, T. Thundat
and R. J. Warmack, Phys. Rev. Lett. {\bf 77}, 1330 (1996).

\bibitem{Searson} P. C. Searson, R. Li and K. Sieradzki, Phys. Rev.
Lett. {\bf 74}, 1395 (1995).

\bibitem{Hwang} R. Q. Hwang, J. C. Hamilton, J. L. Stevens and S. M.
Foiles, Phys. Rev. Lett. {\bf 75}, 4242 (1995).

\bibitem{RajAshby} R. Raj and M. F. Ashby, Metall. Trans. {\bf 2},
1113 (1971).

\bibitem{Wei} Y. J. Wei and L. Anand, J. Mech. Phys. Sol. {\bf 52},
2587 (2004).

\bibitem{Vetrano1} J. S. Vetrano, E. P. Simonen and S. M. Bruemmer,
Acta Mater. {\bf 47}, 4125 (1999).

\bibitem{Vetrano2} J. S. Vetrano, C. H. Henager and E. P. Simonen, in
{\it Superplasticity -- Current Status and Future Potential}, edited
by P. B. Berbon, M. Z. Berbon, T. Sakuma and T. Langdon. (Materials
Research Society, Warrendale, PA, 2000).

\bibitem{Hamilton1} C. H. Hamilton, C. C. Bampton and N. E. Paton in
{\it Superplastic Forming in Structural Alloys}, edited by N. E. Paton
and C. H. Hamilton. (The Metallurgical Society of AIME, Warrendale,
PA, 1982).

\bibitem{Pilling1} J. Pilling and N. Ridley, in {\it Aluminium
Technology '86}, edited by T. Sheppard. (Institute of Metals, London,
UK, 1986)

\bibitem{Cope1} M. T. Cope, M. Sci. Thesis, Victoria University of
Manchester, U. K. (1982).

\bibitem{Kaibyshev} R. Kaibyshev, F. Musin, D. R. Lesuer and T. G.
Nieh, Mater. Sci. and Eng. A {\bf 342}, 169 (2003). 

\bibitem{Vanderhasten1} M. Vanderhasten, L. Rabet and B. Verlinden, J.
Mat. Eng. Performance {\bf 16}, 208 (2007).

\bibitem{Vanderhasten2} M. Vanderhasten, L. Rabet and B. Verlinden,
Materials and Design {\bf 29}, 1090 (2008).

\bibitem{Song} R. Song, D. Ponge and D. Raabe, Acta Mater. {\bf 53},
4881 (2005).

\bibitem{Hayes}	J. S. Hayes, R. Keyte and P. B. Prangnell, Mater. Sci.
and Technol. {\bf 16}, 1259 (2000).

\bibitem{Wang}	Z. C. Wang and P. B. Prangnell, Mater. Sci. and Eng. A
{\bf 328}, 87 (2002).

\bibitem{Hutanu} R. Hutanu, L. Clapham and R. B. Rogge, Acta Mater.
{\bf 53}, 3517 (2005).



\end{thebibliography}
\end{document}